\documentclass{elsart4-1}

\usepackage{graphicx}
\usepackage{epsfig}
\usepackage{xspace}
\usepackage{url}

\usepackage{amssymb}

\usepackage[english,francais]{babel}

\newcommand{\g}{$\gamma$\xspace}
\newcommand{\rhoGJ}{$\rho_{GJ}$\xspace}
\newcommand{\Pdot}{$\dot{P}$\xspace}
\newcommand{\Edot}{$\dot{E}_{rot}$\xspace}

\newcommand{\epair}{$e^-$-$e^+$\xspace}

\newcommand{\degr}{$^{\circ}$\xspace}

\newcommand{\Fermi}{\textit{Fermi}\xspace}

\def\gsim{\mathrel{\raise.5ex\hbox{$>$}\mkern-14mu
             \lower0.6ex\hbox{$\sim$}}}
\def\lsim{\mathrel{\raise.3ex\hbox{$<$}\mkern-14mu
             \lower0.6ex\hbox{$\sim$}}}
\def\be{\begin{equation}}
\def\ee{\end{equation}}
\def\og{\leavevmode\raise.3ex\hbox{$\scriptscriptstyle\langle\!\langle$~}}
\def\fg{\leavevmode\raise.3ex\hbox{~$\!\scriptscriptstyle\,\rangle\!\rangle$}}

\begin{document}

\centerline{Astrophysics}
\begin{frontmatter}



\selectlanguage{english}
\title{Gamma-ray pulsars: a gold mine}


\selectlanguage{english}
\author[authorlabel1]{Isabelle A. Grenier}
\ead{isabelle.grenier@cea.fr}
\author[authorlabel2]{Alice K. Harding}
\ead{Alice.K.Harding@nasa.gov}

\address[authorlabel1]{Laboratoire AIM Paris-Saclay, CEA/Irfu, CNRS, Universit\'e Paris Diderot, 91191 Gif-sur-Yvette cedex, France.}
\address[authorlabel2]{NASA Goddard Space Flight Center, Greenbelt, MD 20771}

\medskip
\begin{center}
{\small Received *****; accepted after revision +++++}
\end{center}

\begin{abstract}
The most energetic neutron stars, powered by their rotation, are capable of producing pulsed radiation from the radio up to \g rays with nearly TeV energies. These pulsars are part of the universe of energetic and powerful particle accelerators, using their uniquely fast rotation and formidable magnetic fields to accelerate particles to ultra-relativistic speed.
The extreme properties of these stars provide an excellent testing ground, beyond Earth experience, for nuclear, gravitational, and quantum-electrodynamical physics.
A wealth of \g-ray pulsars has recently been discovered with the Fermi Gamma-Ray Space Telescope. The energetic \g rays enable us to probe the magnetospheres of neutron stars and particle acceleration in this exotic environment.  
We review the latest developments in this field, beginning with a brief overview of the properties and mysteries of rotation-powered pulsars, and then discussing \g-ray observations and magnetospheric models in more detail.

\vskip 0.5\baselineskip
\selectlanguage{francais}
\noindent{\bf R\'esum\'e}
\vskip 0.5\baselineskip
\noindent
{\bf Les pulsars \g: une mine d'or:}
Les \'etoiles \`a neutrons les plus puissantes, qui tirent leur \' energie de leur rotation, sont capables d'\' emettre des impulsions lumineuses des ondes radio jusqu'aux rayons \g d'\' energies proches du TeV. Ces pulsars font partie des puissants acc\' el\' erateurs de particules de l'Univers, profitant de leur rotation unique et de leur formidable champ magn\' etique pour acc\'el\'erer des particules jusqu'\`a des vitesses ultra-relativistes. Les propri\' et\' es extr\^emes de ces \'etoiles permettent de tester la physique nucl\' eaire, la gravitation et l'\' electrodynamique quantique dans des conditions inaccessibles sur Terre. Le t\' elescope gamma spatial \Fermi vient de r\' ev\' eler un riche \' echantillon de pulsars \g. Leur rayonnement \g \' energ\' etique permet de sonder la magn\' etosph\`ere des \' etoiles \`a neutrons et d'\' etudier l'acc\' el\' eration de particules dans cet environnement exotique. Nous pr\' esentons les derniers d\' eveloppements de ce domaine, en commen\c{c}ant par une rapide revue des propri\' et\' es et myst\`eres soulev\' es par ces pulsars, puis en d\' etaillant plus avant les observations \g et  les mod\`eles magn\' etosph\' eriques.

\vskip 0.5\baselineskip
\noindent{\small{\it Keywords~:} pulsar~; neutron star~; magnetosphere~; gamma rays~; acceleration 
\vskip 0.5\baselineskip
\noindent{\small{\it Mots-cl\'es~:} pulsar~; \' etoile \`a neutrons~; magn\' etosph\`ere~; rayons gamma~; acc\'el\'eration}}

\end{abstract}
\end{frontmatter}


\selectlanguage{english}

\section{Neutron stars, pulsars, and their puzzles}
\label{sec:intro}
\selectlanguage{english}
\begin{figure}[!t]
\centering
\includegraphics[width=0.8\textwidth]{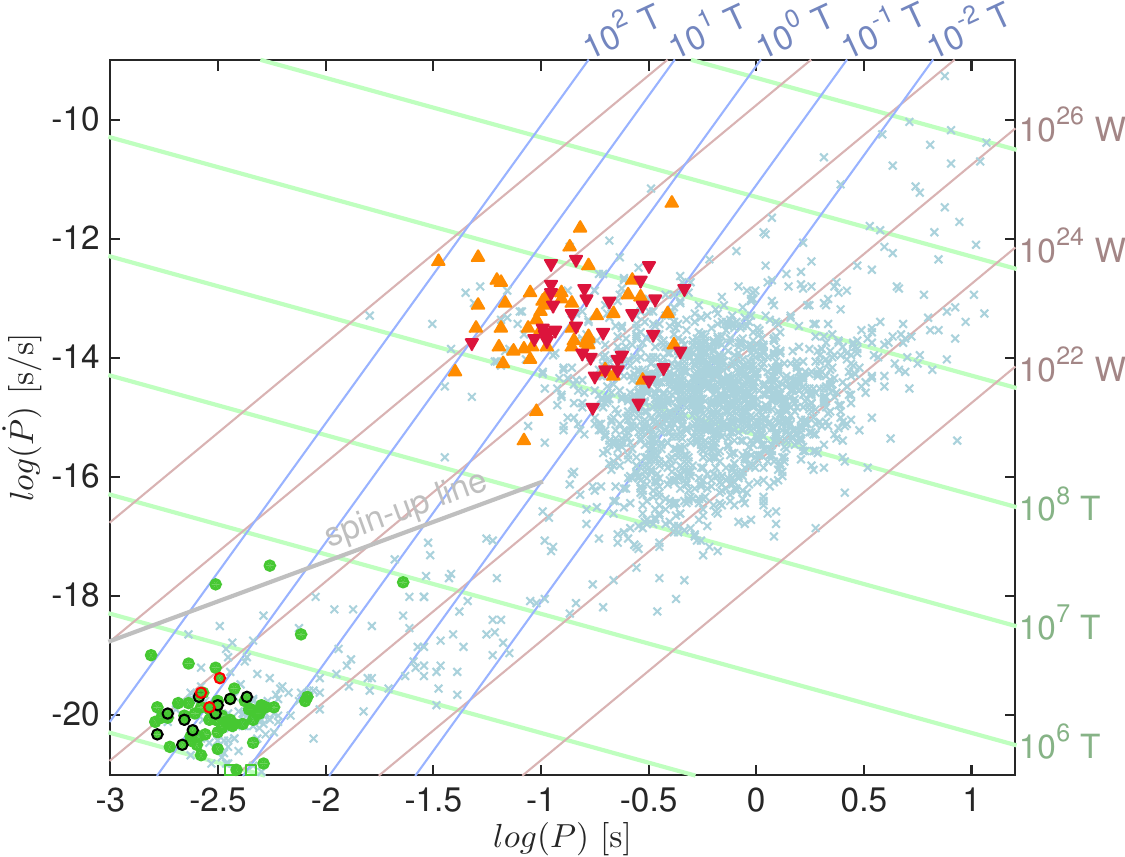}
\caption{Distribution in period (P) and period time derivative (\Pdot) of \g-ray pulsars \cite{LAT_detected_pulsars} and radio pulsars \cite{ATNF} with 53 radio-loud and \g-loud young pulsars (orange upward triangles), 37 radio-faint and \g-loud young pulsars (red downward triangles), 71 radio-loud and \g-loud millisecond pulsars (green filled circles, circled in black and red when in black-widow and redback systems, respectively), and 2256 other radio pulsars (light blue). Recently discovered millisecond pulsars, with no \Pdot measurement yet, are plotted as squares at \Pdot near $10^{-21}$. Lines of constant spin-down power (brown) and polar magnetic field strength (green) are given for a magnetic dipole in vacuum and a stellar moment of inertia of $1.4\,{\times}\,10^{38}$~kg m$^{-2}$ applicable to a 1.4 solar mass neutron star with a 12 km radius \cite{Lattimer07}. Lines of constant magnetic field strength at the light cylinder radius are shown in blue. The grey line marks the spin-up rate expected from mass transfer at the Eddington rate from a stellar companion in a binary system (see equation~\ref{eq:spinup}).}
\label{fig:PPdot}
\end{figure}

A neutron star is an extreme object in many regards. It is a compact, rapidly spinning, highly magnetized star, formed from the core of a massive star which runs out of nuclear power and collapses under its own gravity, while the outer stellar layers rebound and explode into a supernova. By conservation of angular momentum, the collapsed neutron star ends up spinning at the mind-boggling rate of tens of revolutions per second (see Figure \ref{fig:PPdot}). With the equivalent of 1 to 2 solar masses compressed in a sphere the size of a city, with 10 to 14 km radii \cite{Kiziltan13,Antoniadis13,Steiner13}, these objects provide a unique test-ground to study ultra-dense matter. To put these numbers into perspective, imagine a pinhead filled with 200,000 tons of matter, crushed near the surface by a a gravitational force $1.3\,{\times}\,10^{11}$ times larger than on Earth. 

\begin{figure}[!t]
\centering
\includegraphics[width=16cm]{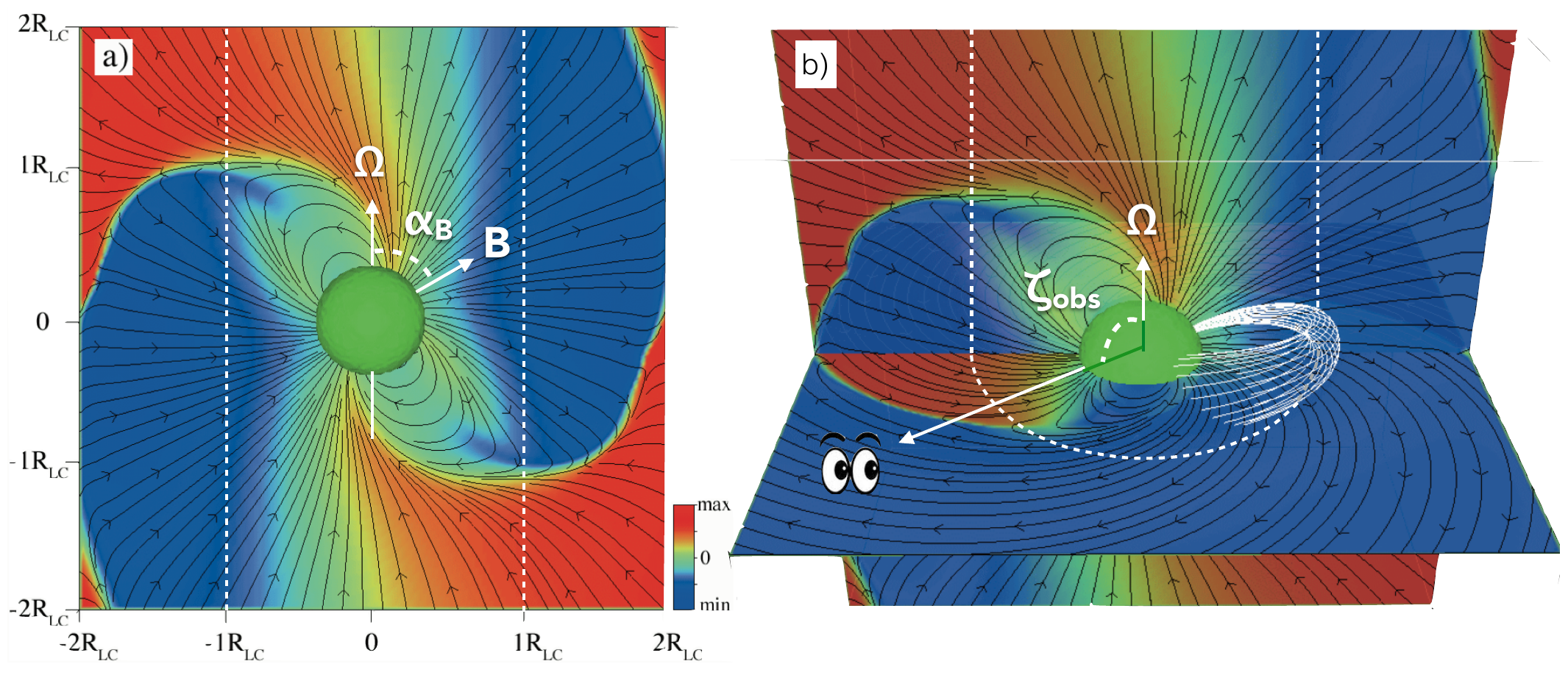}
\caption{Oblique force-free pulsar magnetosphere obtained from MHD simulations \cite{Spitkovsky06} for a magnetic obliquity $\alpha_B = 60^\circ$, showing magnetic fields lines in (a) the plane of the spin and magnetic axes, and (b) slices through the plane of the spin axis and spin equator. Color indicates the magnetic field component perpendicular to the plane. An observer, at a viewing angle $\zeta_{obs}$ to the spin axis, sees the magnetic field corotate with the star near the surface, sweep back near the light cylinder radius (dashed lines), and form a wind transporting fields and plasma outward at larger distance. Field lines that close within the light cylinder define the closed field region and field lines that extend beyond the light cylinder define the open field region.  The boundary of this region at the neutron star surface defines the polar cap.  A current sheet forms where the magnetic field reverses direction near and past the light cylinder (the boundary between the red and blue regions).}
\label{fig:FF}
\end{figure}

\textbf{The magnetic field}, entrained and amplified (and partly generated?) in the birth collapse, is also beyond terrestrial measure. Assuming magnetic dipole radiation we infer polar strengths of $10^8$ to $10^{11}$~T, which correspond to the strongest magnetic fields in the known universe (see Figure \ref{fig:PPdot}). In comparison we measure ${\sim}\,60\,\mu$T near the Earth poles and can briefly reach $10^3$~T in terrestrial laboratories. 
The rotation of the magnetic field, anchored in the star and inclined at an angle $\alpha_B$ to the spin axis (see Figure \ref{fig:FF}), induces a net radial Poynting flux which carries energy away from the star at the expense of its rotational energy. 
The young neutron stars thus slow down over a few $ 10^8$~years, from initial spin periods $P$ of 30-100~ms and rotational powers \Edot of $10^{30-32}$~W, down to periods of order 1 s and to powers as low as $10^{22-24}$~W (see Figure \ref{fig:PPdot}). In comparison, the total radiation yield from the Sun represents $3.8\,{\times}10^{26}$~W.
In the course of their evolution, some neutron stars born in binary systems may spin up by accreting angular momentum from a low-mass star companion, if the geometry of the mass transfer is favourable. In Figure \ref{fig:PPdot}, these recycled pulsars ``drift'' below the maximum spin-up line, up to periods of a few milliseconds, hence their name of millisecond pulsars (MSP) and their large rotational powers of $10^{26-30}$~W despite their few billion years of age. They have low magnetic fields of $10^{4-6}$~T.

\textbf{The interior structure of a neutron star} defies our understanding of the states of matter at different depths (from the solid ionic crust to neutron matter, to hyperon-dominated matter, condensed mesons, or strange quarks) and of the coupling of the magnetic flux tubes (generated by superconducting protons) with the numerous rotational vortices of the neutron superfluid. The moment of inertia also serves to test gravity theories beyond General Relativity. Mass-radius measurements and observations of the rotational properties and starquake `glitches', and of the thermal emission and long-term stellar cooling, provide complementary constraints on these questions. A few results are discussed in section~\ref{sec:MSPlab} and we point the reader to \cite{Watts15} for further literature.

\begin{figure}[!t]
\centering
\includegraphics[width=16cm]{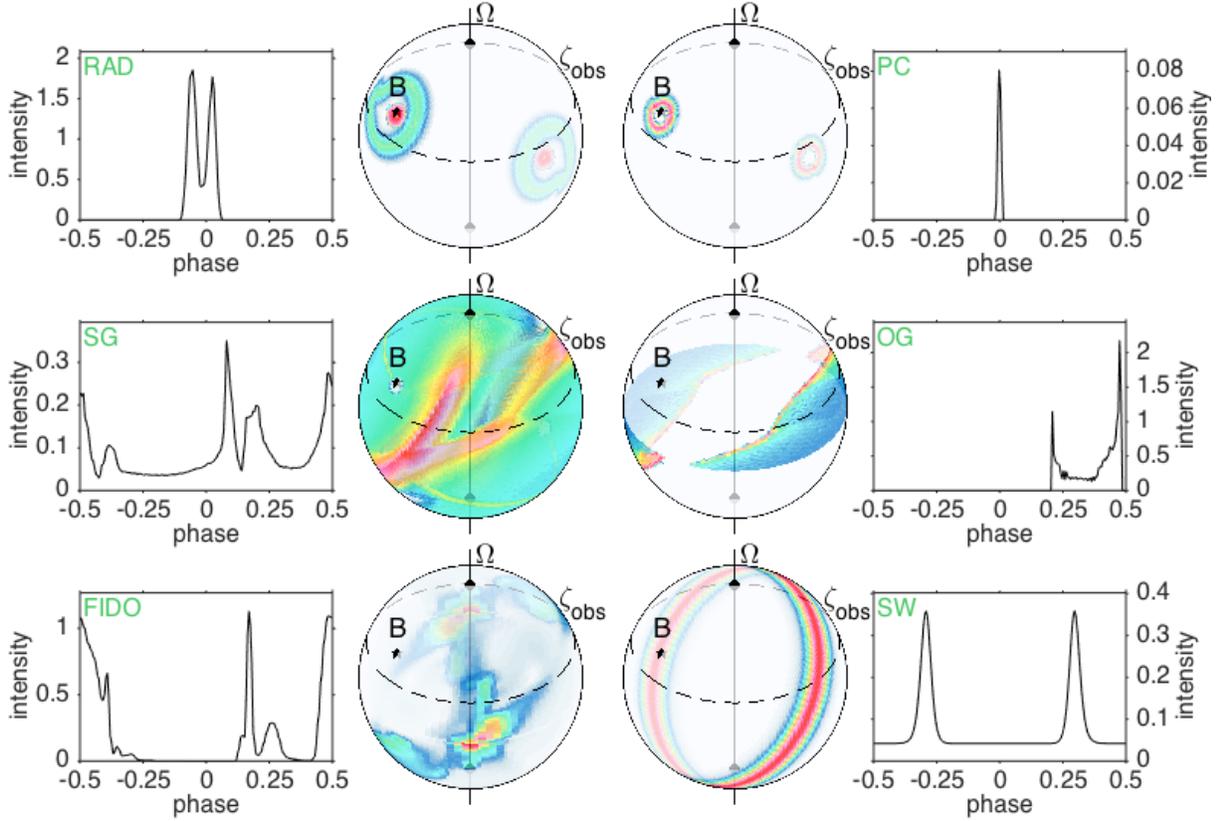}
\caption{Model predictions of the radiation pattern produced across the sky by a 0.1~s period pulsar with a $10^8$~T magnetic field. The magnetic axis (B labeled star) is inclined at 55\degr to the spin axis (vertical line labeled $\Omega$). Intensities increase from blue to red and are normalized to a total of 1 over the whole sky for each model. An observer inclined at  $\zeta_{obs} = 68.5$~\degr to the spin axis (dashed curve) would intercept the corresponding lightcurves. The radio model (RAD) includes a core beam centred on the magnetic axis, surrounded by a hollow cone of emission. Both components are produced at a few hundred kilometres above the polar caps. The \g-ray models compare different emitting regions. Emission arising from a few stellar radii above the polar caps (PC) is focused into hollow cones about the magnetic axis. The slot-gap (SG) and outer-gap (OG) beams are caustics formed near, but inside, the light cylinder, by relativistic aberration and retardation that bunch the radiation along the trailing edge of the open magnetosphere \cite{Dyks04}. The slot gap extends from near the star to the light cylinder whereas the outer gap is restricted to high altitudes. The FIDO and striped-wind (SW) beams are caustics formed outside the light cylinder by stagnation of outflowing, radiating particles, on sweptback field lines \cite{BS10}. Emission is produced slightly beyond the light cylinder in the FIDO case (Force-free Inside the light cylinder and Dissipative Outside) and at 5 times the light-cylinder radius in the striped wind. The latter bears this name because the magnetic field configuration far from the star, in the wave zone, consists of stripes of alternating toroidal field \cite{Coroniti90}} 
\label{fig:phaseplots}
\end{figure}

\textbf{Neutron stars have been dubbed pulsars} when found to emit pulses of radiation in the radio in 1967 \cite{Hewish68}, in the optical and X rays in 1969 \cite{Cocke69,Fritz69,Bradt69}, and in \g rays in the early 70s \cite{Kniffen74}. Pulses of short duration and stable frequency were rapidly explained as rotating light beams sweeping past the Earth at the angular velocity, $\Omega = 2 \pi/P$, of the star : only dense stars such as neutron stars could spin fast enough to match the observed frequencies \cite{Gold68}. Since then, over 2400 pulsars have been discovered across the sky at different wavelengths (see Figure~\ref{fig:PPdot}). The narrow, lighthouse-like, beams that were first conceptualized to explain the radio pulses have evolved into a variety of shapes according to the radiation wavelength and its origin in the pulsar magnetosphere, whether the emission is produced at low altitude above the star or further out near the light cylinder. This cylinder marks the boundary, at radius $R_{LC}\,{=}\,c/\Omega$ from the spin axis, beyond which the field lines cannot co-rotate with the star, but start to spiral outward (see Figure~\ref{fig:FF}). The region of field lines that do not close within the light cylinder is often referred to as the open magnetosphere and the foot points of these lines on the stellar surface define the polar caps. Figure~\ref{fig:phaseplots} shows that beam apertures can vary from hollow flaring cones, when generated at tens of stellar radii above the magnetic polar caps, to fan-like beams when produced near and beyond the light cylinder. So far, all invoked radiation sites relate to the open magnetosphere or to the undulating equatorial current sheet, near or past the light cylinder (see Figure~\ref{fig:phaseplots}), but the provenance and shape of the radiation beams remain uncertain (see section~\ref{sec:mag}). 

\textbf{Neutron stars can develop rich magnetospheres}, filled with dense plasma, because of the large electric fields induced by the rotation of the magnetic field anchored to the stellar crust. Faraday would have revelled in the study of these extreme unipolar inductors, which can generate an EMF of order $10^{17}$~V (for a young, $P\,{\sim}0.1$~s, $B\,{\sim}10^8$~T star). The electric fields can overcome the intense gravity to fill the magnetosphere with charges \cite{Goldreich69}.  Copious production of electron-positron pairs from primary \g rays could readily fill at least parts of the magnetosphere with the so-called Goldreich-Julian (GJ) charge density, \rhoGJ, that is required to achieve near force-free conditions ($\rho \overrightarrow{E} + \overrightarrow{j} \times \overrightarrow{B} = 0$) and co-rotation of the magnetosphere with the star. It took decades to numerically find the force-free configuration of a charge-filled magnetosphere for an inclined magnetic moment (see section~\ref{sec:mag}). But the models cannot yet self-consistently explain the origin or distribution of the charges, nor predict whether the large GJ densities can be met in reality to achieve force-free conditions or, a contrario, leave regions of net acceleration (see section~\ref{sec:mag}).

\textbf{Particle acceleration} along the magnetic field lines takes place where the charge density differs from the local value of \rhoGJ ($\overrightarrow{\nabla} \cdot \overrightarrow{E}_{\parallel} = (\rho - \rho_{GJ})/\epsilon_0$). The  $E_{\parallel}$ field is strong enough to boost the particles to relativistic energies and the guiding geometry of the magnetic field ensures focussed beams of radiation.
The curved field lines impose a perpendicular acceleration which causes the accelerated particles to emit primary curvature radiation at \g-ray energies. The potential drop available for particle acceleration is limited by the subsequent development of electron-positron pair cascades which can bring the charge density back to \rhoGJ and screen the accelerating electric field $E_{\parallel}$. Near the neutron star, \g-ray photons convert to \epair pairs in the strong magnetic field. Conversely, in the outer magnetosphere, pair cascades are triggered by two-photon interactions between \g rays and low-energy photons, in particular thermal X rays emerging from the neutron-star surface and/or polar caps heated by the return current.
For large pair multiplicities, the efficient screening leads to thin acceleration zones (gaps) and to efficient saturation of the particle Lorentz factors, with values decreasing from $10^8$ to $10^6$ with the global conductivity of the magnetosphere \cite{Kala14}. Conversely, in pair-starved conditions, the accelerating region partially or entirely fills the open magnetosphere. 
This feedback between acceleration, radiation, and cascading affects the emission geometry, but also the total \g-ray yield from a pulsar. The full voltage across the open field lines evolves as $V_0 = 6 \times 10^{12} \rm{V}\,(P/1\,\rm{s})^{-2}\,(B/10^8 \rm{T})$. It exceeds $10^{15}$ V for the most energetic pulsars, but the efficient cascading and screening induced by particles with energies $E_p \sim 10^{13}$ eV limit further acceleration. The number rate of these primary particles and their energy $E_p$ imply a \g-ray luminosity $L_\gamma \propto \dot{E}_{rot}^{1/2}$ \cite{Harding81,Harding02}. Less energetic pulsars having $V _0 < 10^{13}$ V do not produce enough pairs to efficiently screen the electric field and they use all of the open field voltage to accelerate particles and produce high-energy radiation. The voltage $V_0$ and number rate of GJ charges yields $L_{\gamma} \propto \dot{E}_{rot}$ in that case. The current sample of young and millisecond \g-ray pulsars exhibit luminosity evolutions close to these two trends, but there is a large scatter and it is unclear how much of the magnetosphere is screened in each object, therefore where are the different dissipative zones and how they contribute to the lightcurve of a given pulsar. In this context, the GeV photons produced by the primary particles are invaluable to pinpoint the location of the acceleration sites since the particles radiate their energy `on the spot'.

\textbf{Coherent radio emission} is produced by bunches of particles at the end of the cascade process. The emission is probably focused into two beams: a core beam close to the magnetic axis, and a surrounding hollow or `conal' beam of period-dependent width and altitude, widening for shorter-period pulsars. Sampling different lines of sight across the beams results in a variety of single or multiple radio pulses \cite{Rankin83}  (see Figure~\ref{fig:phaseplots}). Complex pulse shapes may also require a ``patchy'' beam with large intensity variations inside the conal beam \cite{Lyne88}. The extremely large brightness temperatures observed in the radio are due to coherent radiation. Because the coherent volume is smaller at shorter wavelengths, most pulsars have steep radio spectra, with energy flux distributions typically close to $S_{\nu} \propto \nu^{-1.7}$ in frequency, but a few extreme slopes below -3 and above -0.5 have been recorded. There is not yet a complete theory of coherent radio emission from pulsars that accounts for the observed phenomena, although leading ideas include curvature radiation from bunched particles due to the two-stream instability \cite{Lyubarski02} or maser curvature or synchrotron emission \cite{Melrose04}.

\textbf{Time variability} in pulsars is also greatly puzzling. The abundant new data stress an apparent disconnection between the long-term stability of the average multi-wavelength pulses, which is suggestive of a stable magnetospheric configuration, against time changes recorded in the neutron-star rotation (glitches) \cite{Espinoza11}, intermittency of the radio emission over many timescales \cite{Keane13}, and mode switches to different, but recurrent, pulse properties in the radio and X rays \cite{Hermsen13}. The correlation with changes in the angular velocity of the star imply unexplained modifications of the current circulation in the magnetosphere and of the braking torque.  A \g-ray mode change has now been added to the list of variability puzzles (see section~\ref{sec:imp}).

\textbf{Pulsar wind nebulae} often envelop pulsars as extended sources of broadband non-thermal emission, seen from radio to X-ray and \g-ray wavelengths. They signal the presence of a magnetized wind of relativistic particles which takes away most of the pulsar rotational power.  Despite recent advances in relativistic simulations, we still do not understand how the pulsar energy is divided between the magnetic dipole radiation and relativistic particles. The Poynting flux dominates near the star, but the particle energy seem to dominate near the wind's termination shock with the ambient medium \cite{KC84,Bucciantini14}. The Crab nebula is the proto-typical example and has long served as an astronomical standard candle, but recently detected \g-ray flaring has shattered the placid image of pulsar wind nebulae (see Section \ref{sec:Crab}).

In this dynamic context, the recent discovery of numerous \g-ray pulsars brings a fresh perspective and wealth of new information on these long standing questions, which we review in more details below.

\begin{figure}[!t]
\centering
\includegraphics[width=0.85\textwidth]{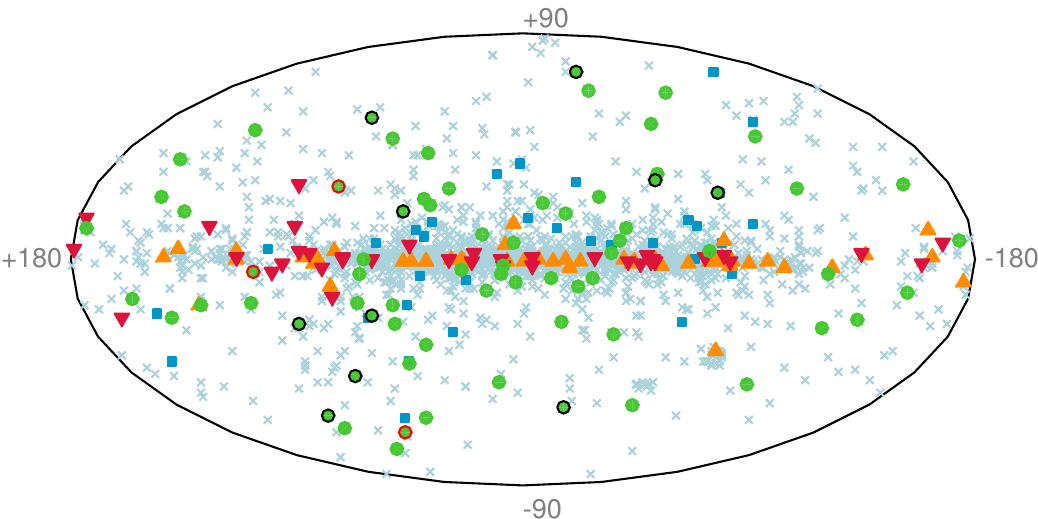}
\caption{Position in Galactic coordinates of the same radio and \g-ray pulsars as in Figure~\ref{fig:PPdot}, with radio+\g-loud young pulsars shown as orange upward triangles, radio-faint and \g-loud young pulsars denoted as red downward triangles, and radio+\g-loud millisecond pulsars shown as green filled circles, circled in black and red when in black-widow and redback systems, respectively. Blue squares highlight the radio pulsars that have been recently discovered in the direction of \Fermi-LAT point sources and which await timing solutions to possibly detect \g-ray pulsations. Other radio pulsars are shown in light blue.}
\label{fig:allsky}
\end{figure}

\section{The \Fermi legacy}
\label{sec:Fermi}

The \Fermi Gamma-ray Space Telescope has been launched into orbit in 2008 and has continuously mapped the sky in \g rays since then \cite{Thompson15}. The satellite carries two instruments, among which the Large Area Telescope (LAT), which has enabled the discovery of 161 \g-ray pulsars, instead of the handful previously known. The number is still increasing as we write \cite{LAT_detected_pulsars}. The LAT is sensitive to \g rays with energies from 20 MeV to over 300 GeV, in a large field-of-view covering 2.4 sr. Most of the time, the satellite rocks in order to image the entire sky every two orbits (${\sim}\,3$ hours). Photons are time-stamped to an accuracy better than 1 $\mu$s, so that rapid pulsations from millisecond pulsars can be retrieved.

\subsection{The \Fermi treasure chest}
\label{sec:Fermichest}

The all-sky survey has proved to be a treasure chest for pulsars. The LAT sensitivity and pulsar \g-ray fluxes are such that we need weeks to months of observations to gather enough photons to detect a pulsation and build the average pulse profile over one stellar rotation (i.e. light curve).  Yet, a pulsar rotates hundreds or thousands of times between two photon detections, even for the Vela pulsar, which is the brightest \g-ray source in the sky. We thus need to know precisely how a pulsar slows down to fold the photons in rotational phase. This requirement is less stringent for MSPs because of their very low period derivatives (see Figure~\ref{fig:PPdot}). Photon arrival times must also be converted to their value at the Solar System barycentre to correct for light travel time differences as the telescope moves about the Earth and Sun.
An international Pulsar Timing Consortium has organized extensive timing campaigns, involving all major radio telescopes and targeted X-ray satellite observations \cite{Smith08}. This impressive and on-going effort has proved very successful and enabled the discovery and follow-up studies of tens of \g-ray pulsars (see Figs.~\ref{fig:PPdot} and \ref{fig:allsky}). Regular monitoring is essential to correct the timing solutions for timing noise and glitches, and for orbital motion when in binary systems.

In parallel, the decade-long quest that led to the identification of the bright radio-quiet Geminga pulsar in the 1980s \cite{Caraveo14} had paved the way to search for \g-ray pulsars with no radio counterpart and only faint thermal X-ray emission, barely visible to a few hundred parsecs. Algorithms were developed before the \Fermi launch to ``blindly'' search for periodicity in \g-ray data. Success was met by analyzing the differences of photon arrival times over time spans of a few weeks, instead of using the computationally prohibitive Fourier analyses \cite{Atwood06}. Tens of radio-quiet \g-ray pulsars have thus been discovered in blind searches (see Figs.~\ref{fig:PPdot} and \ref{fig:allsky}), starting with the identification of J0007+7303 in the CTA 1 supernova remnant \cite{Abdo08_CTA1}. This neutron star is one of the few \g-ray pulsars with a polar magnetic field in excess of $10^9$~T (see Figure~\ref{fig:PPdot}). Methods have also been developed to use the volunteer supercomputer Einstein@Home. Volunteer citizens, spread around the Earth, have thus contributed to the discovery of four \g-ray pulsars \cite{Einsteinathome}. 

A vivid account of these discoveries and their references are given in \cite{Caraveo14}. Whereas the number of new young \g-ray pulsars fell close to expectations from various models \cite{Pierbattista12}, the large number of MSP identifications came as a surprise, as discussed in section~\ref{sec:oldPSR}.

\subsection{The \Fermi treasure map}
\label{sec:Fermimap}

Besides revealing \g-ray pulsars by folding the \g-ray data with known ephemerides, the \Fermi-LAT survey has been used as a treasure map to discover neutron stars in the direction of suitable \g-ray sources drawn from the \Fermi catalogues of point sources \cite{Abdo09_BSL,Abdo10_cat1,Nolan12,Acero15_3FGL}. In particular, it was suggested that deep radio searches of \Fermi-LAT sources could discover MSPs, leading to detection of their \g-ray pulsations \cite{Story07}.
A Pulsar Search Consortium (PSC) undertook searches at radio and X-ray wavelengths at the positions of unidentified \g-ray sources that were non-variable and had hard spectra with high-energy cutoffs.
In 2012, the consortium had already found 43 MSP and 4 young pulsars out of 300 \g-ray sources \cite{Ray12}. This is a phenomenal success rate and showed that \Fermi could pinpoint the locations of numerous new Galactic MSPs.
The timing solutions were then applied to the \Fermi-LAT data to find the \g-ray pulsation. This radio-\g-ray synergy works extremely well since virtually all the new radio MSPs showed \g-ray pulsations. It greatly helps since 80\% of them are in binary systems where the pulse arrival times are strongly influenced by the orbital motion.  The reverse situation is interestingly less efficient: only ${\sim}\,10$\% of the pulsars discovered in blind frequency searches of the \Fermi-LAT data show radio pulsations.

To date, 74 new pulsars have been discovered by the PSC and other groups toward \Fermi-LAT sources, 70 being MSPs (29 radio and 41 radio$+\gamma$), and 4 young pulsars (3 radio and 1 radio$+\gamma$).  As evident in Figure~\ref{fig:allsky}, the new MSPs are widely distributed throughout the sky as they have moved over billions of years from their birthplaces in the Galactic plane. Since only around 60 MSPs were known in the Galaxy outside of globular clusters before \Fermi launch, \Fermi has significantly contributed to the recent increase in this population, which now numbers over 200.   
\Fermi-LAT has also detected \g rays from several globular clusters \cite{Abdo09_GC,Abdo10_GC,Kong10} that are known to host large numbers of MSPs.  It was conjectured that the \g-ray flux be due to the combined pulsed emission from their MSPs.  This was confirmed when the pulsations from J1823-3021A in NGC6624 were discovered \cite{Freire11}. The flux from this one pulsar accounts for the total cluster flux.

The PSC searches also discovered a high fraction of ``Black Widow" and ``Redback" systems, where an energetic MSPs is ablating its companion star in a tight binary system (see Section \ref{sec:BW}).

\begin{figure}[!t]
\centering
\includegraphics[width=\textwidth]{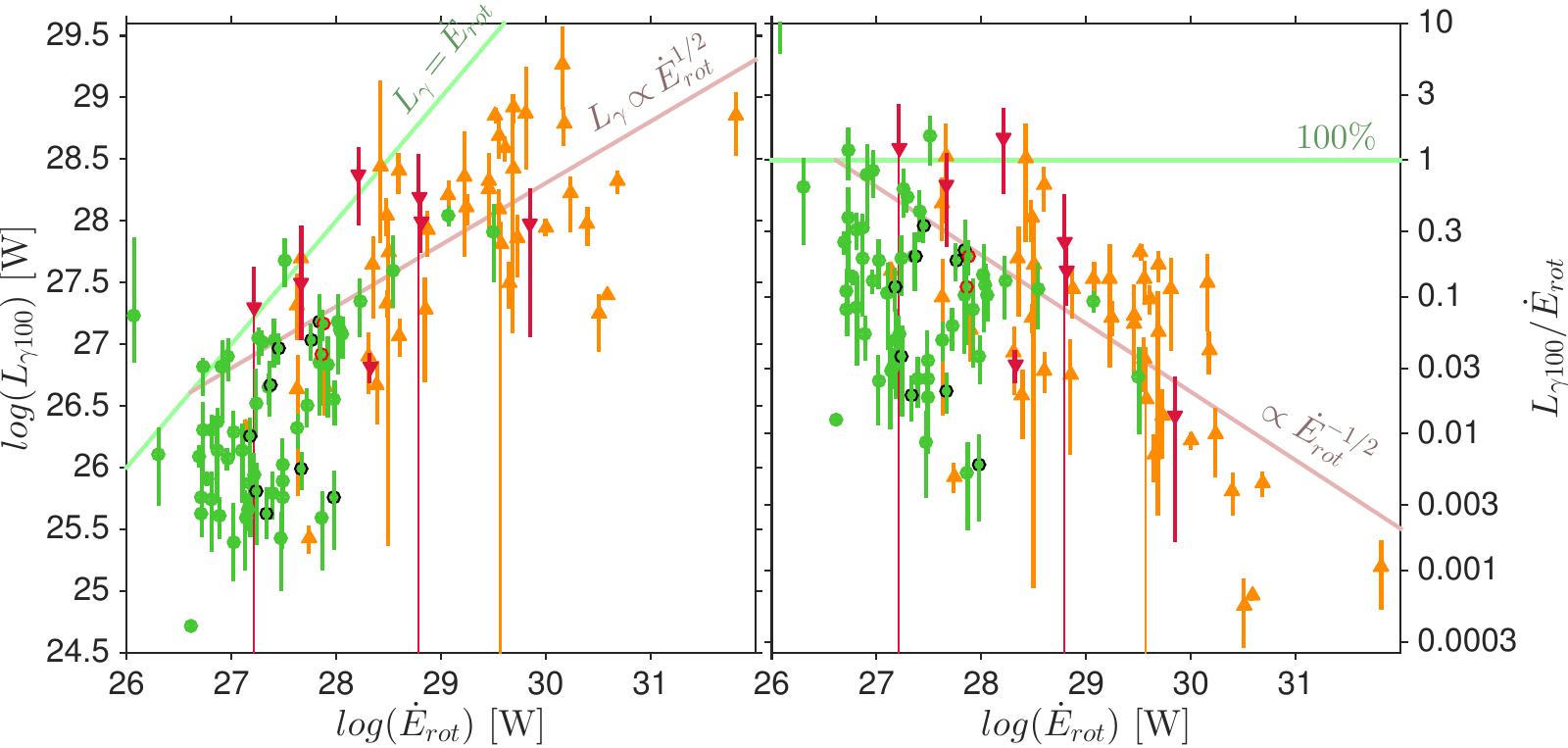}
\caption{\g-ray luminosity above 100 MeV (left) and \g-ray efficiency (right) as a function of Shklovskii-corrected rotational power \Edot, for the subset of \g-ray pulsars of Figure~\ref{fig:PPdot} with distance estimates from \cite{Abdo13_2PC} or inferred from dispersion measures with 30\% error \cite{ATNF}. The error bars include the error on the flux and distance. The green line marks the 100\% efficiency limit and the brown one shows the best $L_{\gamma} \propto \sqrt{\dot{E}_{rot}}$ fit to the young (not MSP) pulsar data. Young pulsars are denoted as orange upward triangles when radio-loud and as red downward triangles when radio-faint. Radio+\g-loud millisecond pulsars are shown as green filled circles, circled in black and red when in black-widow and redback systems, respectively.}
\label{fig:lum_effi}
\end{figure}

\section{From six to ninety young \g-ray pulsars}
\label{sec:youngPSR}

The sample of young \g-ray pulsars has dramatically increased, by more than tenfold, with \Fermi. The first result has settled a two-decade-long debate on the origin of the \g rays in the inner or outer parts of a magnetosphere \cite{Abdo09_Vela}. Magnetic pair production occurs in the strong fields present above the polar caps. The field strength has to satisfy the relation $B \sin \theta > 4.5 \times 10^7$~T to absorb a photon with energy above 100 MeV and pitch angle $\theta$ relative to the field direction. Close to the stellar surface, pair conversion thus yields steep, super-exponential, absorption cut-offs in the \g-ray spectra above a few GeV, at variance with the observations of bright young pulsars which show that the cut-offs are more gradual than exponential, therefore that the \g rays are produced beyond a few stellar radii above the stellar surface. The presence of strong caustic peaks in numerous \g-ray lightcurves further points to \g-ray production at high altitude in the magnetosphere, near the light cylinder or slightly beyond it \cite{Abdo13_2PC}. The comparison of the young and millisecond pulsars in Figure~\ref{fig:PPdot} also stresses the importance of the magnetic field strength near the light cylinder to detect the \g-ray pulsations given the LAT sensitivity.

\g rays dominate the radiative output of young pulsars. The photon spectra are hard, with power-law distributions clustering near $E^{-1.5 \pm 0.4}$ below cut-off energies around $2.9\,{\pm}\,2.0$ GeV. Yet, pulsed emission extends to much higher energies in the brightest pulsars ($> 79$~GeV for Vela \cite{Leung14}, and to $400$~GeV for the Crab \cite{Aleksic12_Crab}). In the Vela case, the highest-energy photons correspond to the gradual decline of the spectrum of the second peak. This sub-exponential cut-off can be conceptually explained by the blending in space of curvature radiation produced with different spectra throughout the caustic region \cite{Abdo10_Vela}, or by the blending in time of curvature radiation produced in outer gaps threaded by variable currents \cite{Leung14}. In the Crab case, the highest-energy photons correspond to an additional radiation component \cite{Du12_annulargap}, produced for instance by secondary and tertiary generations of pairs near the light cylinder, up-scattering the IR-UV synchrotron photons of the cascades \cite{Aleksic12_Crab}.  

Paradoxically, the young pulsars become more efficient at producing \g rays as they slow down and lose less rotational power (see the orange and red data in Figure~\ref{fig:lum_effi}).
\g-ray luminosities, $L_{\gamma}$, have been derived from the observed energy flux ($S_{\gamma}$) when a distance estimate ($D$) is available. The derivation assumes a unique beaming factor, $f_{\Omega}\,{=}\,\frac{L_{\gamma}}{4 \pi D^2 S_{\gamma} }\,{\approx}\,1$, which corresponds to a fan-like beam produced in the outer magnetosphere \cite{Watters09,Pierbattista12}.
A negligible fraction (${\sim}\,0.1$\%) of \Edot is radiated away above 100 MeV at an early age, but this fraction rises to 10--30\% after $10^4$ to $10^6$ years. A luminosity trend close to $L_{\gamma}\,{\propto}\,\sqrt{\dot{E}_{rot}}$ is observed. It suggests that a large part of the magnetosphere is force-free and that \g rays are produced in thin gaps with efficient pair feedback on the voltage drop. The large dispersion about this trend, by a factor of four on each side, may reflect intrinsic differences in gap width (voltage drop), but also extrensic geometrical differences in magnetic obliquities and viewing angles, thus in $f_{\Omega}$. Beaming factors can, in particular, vary by one or two orders of magnitude for radio-quiet objects  \cite{Pierbattista12}. Distance estimates also differ in reliability \cite{Abdo13_2PC}.

\begin{figure}[!t]
  \centering
   \includegraphics[width=16cm]{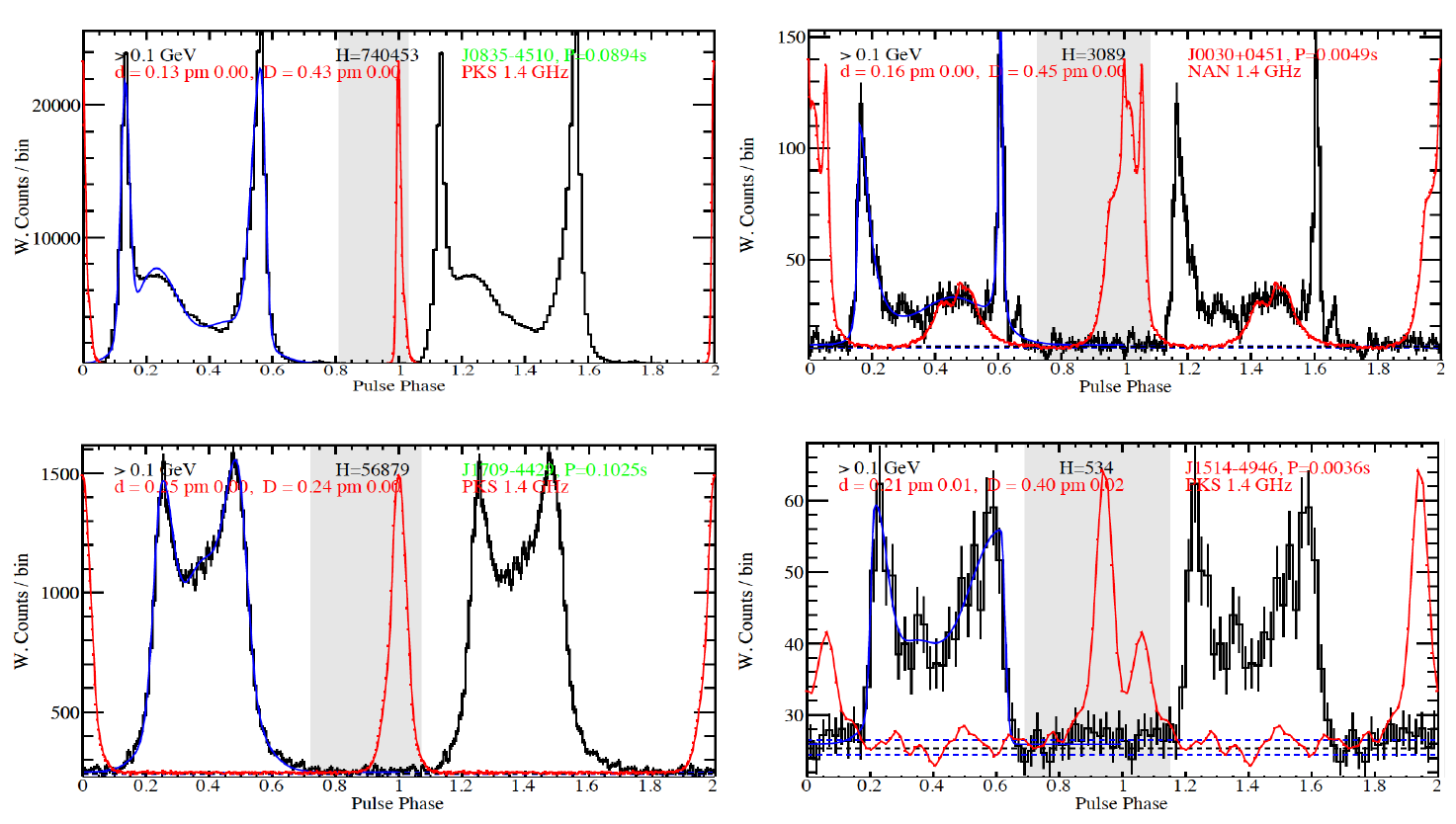}
     \caption{Comparison of \Fermi $\gamma$-ray (black) and radio (red) light curves of young (left) and millisecond (right) pulsars. The young pulsars are J0835-4510 alias Vela (upper) and J1709-4429 (lower). The millisecond pulsars are J0030+0451 (upper) and J1514-4946 (lower), from \cite{Abdo13_2PC}.}
  \label{fig:MSP_LCs}
\end{figure}

Populations of \g-ray pulsars have been simulated for different models in order to disentangle intrinsic trends in the \g-ray yield from complex observational biases (e.g. \cite{Watters11,Takata11_psr,Pierbattista12}). The expectations in total pulsar counts and in the share between \g-only and $\gamma+$radio pulsars agree reasonably well with the data for the outer-magnetospheric models (slot gap and outer gap), thereby corroborating the concept of a much wider beam in \g rays than in the radio.  Yet, these studies stress important shortcomings in the electrodynamics of these models, which predict that $L_{\gamma}\,{\propto}\,w^3 \dot{E}_{rot}$ for a gap width $w$ across the magnetic field lines: (i) the thin gaps that are required to match the sharp \g-ray pulses do not provide enough luminosity to match the observed \g-ray flux, but fall short by two orders of magnitude \cite{Pierbattista14,Hirotani15}; (ii) the models significantly under-predict the abundance of bright \g-ray pulsars with ages less than 100 kyr  \cite{Pierbattista12}. Preliminary studies with the striped wind model also find too few bright \g-ray pulsars despite a different electrodynamical and radiation scheme (Doppler boosted synchrotron radiation from the GJ current heated by reconnection in the current sheet \cite{Petri12}, see section \ref{sec:sheet} and Figure \ref{fig:phaseplots}). This discrepancy cannot be ascribed to visibility uncertainties since the youngest, thus brightest objects remain easily detectable by \Fermi to large distances.  
These problems may hint at the importance of the non-uniform azimuthal distribution of $E_{\parallel}$ electric field and currents around the edge of the open field region, which should concentrate the emission more toward specific directions, thus reducing $f_{\Omega}$. Enhanced $E_{\parallel}$ values from off-centred magnetic fields and polar caps is another possibility. The concentration of emission from near the current sheet may also partly alleviate the luminosity problem (compare the slot-gap and FIDO peak intensities in Figure~\ref{fig:phaseplots}).

\begin{figure}[!t]
  \centering
  \includegraphics[width=0.49\textwidth]{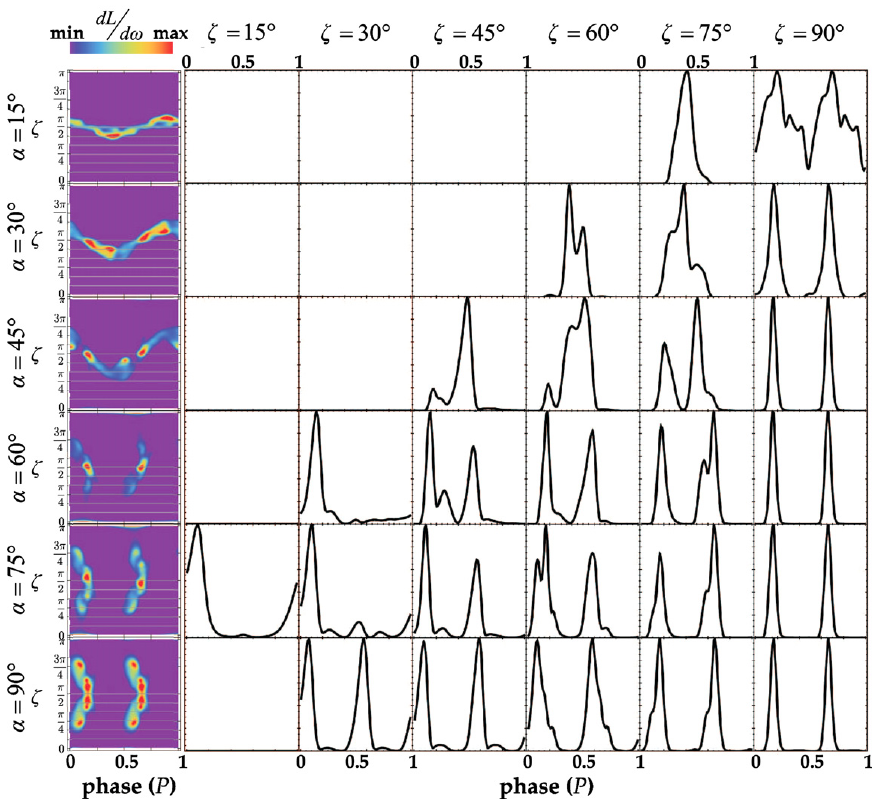} \includegraphics[width=0.49\textwidth]{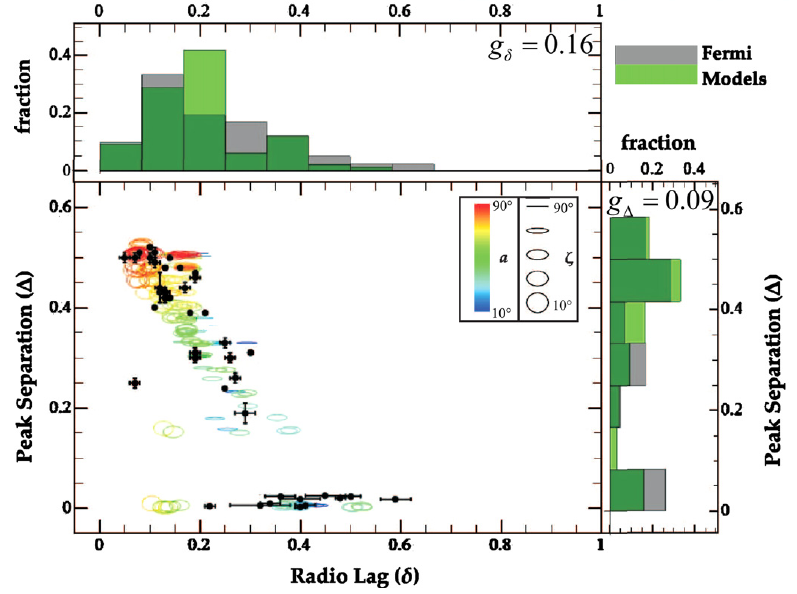}
  \caption{\textit{Left:} skymaps of high-energy luminosity per solid angle $dL/d\omega$ at different inclination angles $\alpha_B$ and light curves at observer angles $\zeta_{obs}$, for a dissipative MHD magnetosphere model, FIDO, having infinite conductivity (force-free) inside the light cylinder and a finite conductivity outside. \textit{Right:} Distribution of peak separations in the \g-ray light-curves versus the phase lag between the radio and \g-ray peaks, for detected \Fermi pulsars (black) and the FIDO model (colored ellipses), from \cite{Kala12b}.}
  \label{fig:FIDO}
\end{figure}

Simulations of dissipative magnetospheres show a clear trend for the emission region to move out from lobes above the polar caps to the current sheet past the light cylinder as one increases the global conductivity of the plasma \cite{Kala14}. This should reflect in the distribution of light curve profiles recorded by \Fermi. Yet, no single geometrical model of the beam origin and shape (see Figure~\ref{fig:phaseplots}) can currently reproduce the whole set of observed light curves \cite{Pierbattista14,Kala14}. Only few trends emerge in the large variety of pulse profiles \cite{Abdo13_2PC}. Intercepting tightly beamed emission above the polar caps is rare \cite{Ravi10}. A majority of profiles  (${\sim}\,3/4$) show two narrow bright peaks, well separated in phase, often bracketing significant bridge emission (see Figure~\ref{fig:MSP_LCs}). The sharp peaks are likely due to the caustic pile-up of photons produced along the edge of the open magnetosphere or near the current sheet. The second (trailing) peak is often harder than the first  \cite{Abdo13_2PC}, which is consistent with curvature radiation. The bridge emission is modulated, up to sometimes forming a third peak, but its origin is unclear. The fact that the third peak of Vela drifts in phase with energy and that it differs in spectrum from the main peaks \cite{Abdo10_Vela} points to a different production site than for the rest of the emission, thereby confirming the possibility of composite \g-ray light curves, intercepting beams produced in different regions \cite{Grenier88}. In the eight luminous pulsars studied in detail so far, the bridge emission appears to be harder and to cut off at earlier energy than the second peak, so its spectral energy distribution (SED) sharply drops about its apex.

\begin{figure}[!t]
\centering
\includegraphics[width=0.5\textwidth]{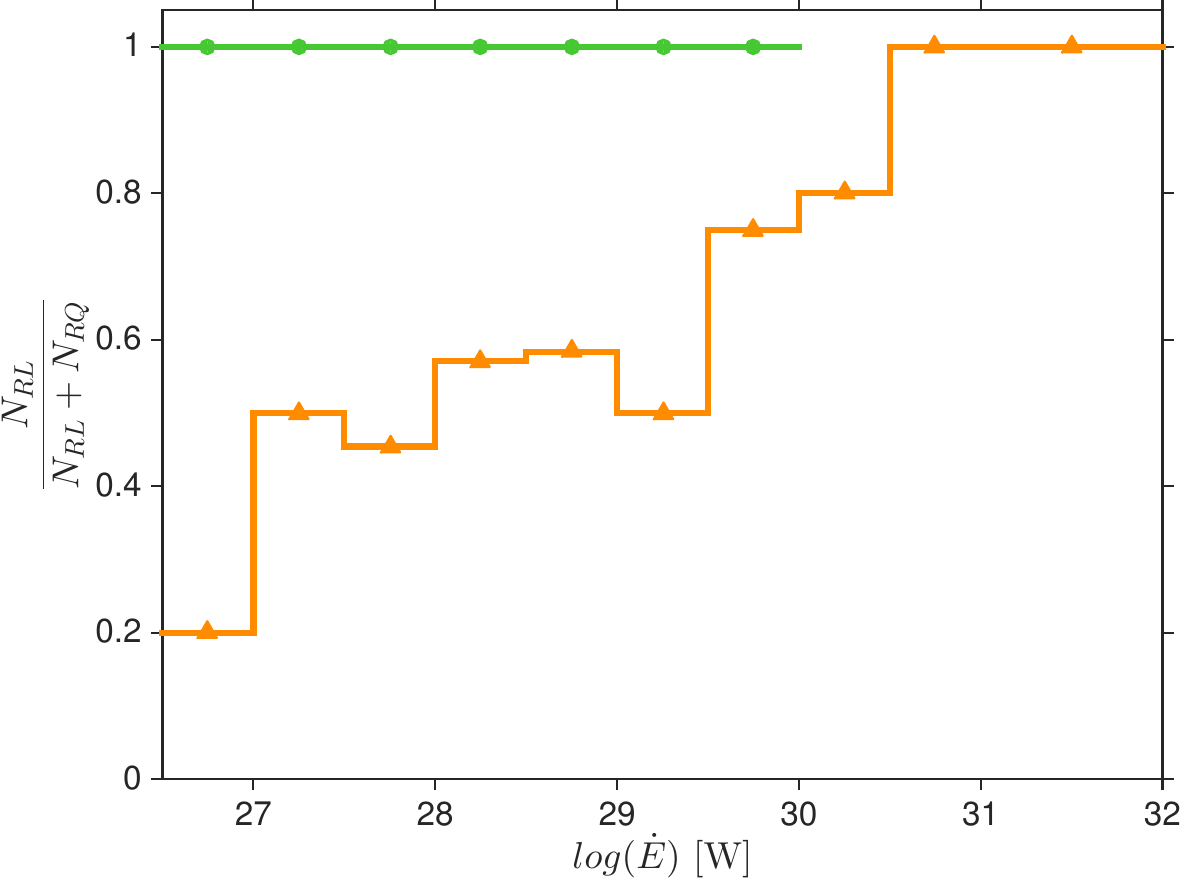}
\caption{Differential evolution with spin-down power of the fraction of radio-detected pulsars in the \g-ray emitting sample, for the young (orange triangles) and millisecond (green circles) pulsars.}
\label{fig:RLfrac}
\end{figure}

The latest magnetospheric simulations show copious pair production near the current sheet for all magnetic obliquities, but not in the open magnetosphere for obliquities below 40\degr \cite{Philippov14}. Model fits to the observed light curves show that the current sample encompasses a large variety of stellar configurations ($\alpha_B$ magnetic obliquity and $\zeta_{obs}$ viewing angle, see Figure \ref{fig:FF}), with low $|\alpha_B - \zeta_{obs}|$ values when we intercept both the radio and \g-ray beams, and large $|\alpha_B - \zeta_{obs}|$ values when we intercept only the wide \g-ray beam \cite{Watters09,Takata11_beam,Pierbattista14}. A useful anti-correlation has been found between the phase separation between the two \g-ray peaks ($\Delta$) and the phase separation between the radio and first \g-ray peak (alias radio lag $\delta$). It is illustrated in Figure~\ref{fig:FIDO}. The formation of caustic \g-ray peaks in the outer magnetosphere can generally explain such an anti-correlation \cite{RY95}, but the small radio lags possibly hint at large plasma conductivities or at emission only near the current sheet \cite{Kala14}, as discussed in section~\ref{sec:mag}. The relative locations of the radio and \g-ray beams remain, however, uncertain since the current models cannot explain the predominance of $\gamma+$radio-loud pulsars at large \Edot power (see Figure~\ref{fig:RLfrac}), whether they use \g-ray emission from one or two poles  \cite{Ravi10,Pierbattista12}, or near the current sheet. This large overlap between the radio and \g-ray beams at high \Edot suggests that the radio emission originates from high-altitudes, near the \g-ray emitting zone, in very young pulsars \cite{Ravi10}

\section{From young to recycled \g-ray pulsars}
\label{sec:oldPSR}
Shortly after the discovery of MSPs in 1982 \cite{Backer82}, an accretion scenario was developed to explain their short periods by transferring angular momentum from a Keplerian accretion disc, fed by a low-mass stellar companion in a binary system \cite{Alpar82}. The neutron star can spin up until its magnetospheric velocity, at the Alfven radius where magnetic pressure balances the infall ram pressure, approaches the Keplerian velocity of the inner edge of the disc. The star reaches the equilibrium period: 
\be
P_{eq} \simeq 0.22\,\rm{ms}\, \left( \frac{\xi}{0.5} \right)^{3/2} \left( \frac{B_{\rm p}}{10^4\,\rm{T}} \right)^{6/7}\,\left( \frac{R_{NS}}{12\,\rm{km}} \right)^{18/7}\,\left( \frac{M_{\rm NS}}{1.4 M_{\odot}} \right)^{-5/7}\,\left( \frac{\dot{M}}{\dot{M}_{\rm Edd}} \right)^{-3/7}
\label{eq:spinup}
\ee
where $B_{\rm p}$, $R_{\rm NS}$, and $M_{\rm NS}$ are the polar magnetic field, radius, and mass of the neutron star; $\xi \sim 0.5$ encapsulates uncertainties in the Alfven radius, and $\dot M_{\rm Edd} \simeq \frac{4 \pi G M_{\rm NS} m_p}{\sigma_T c}$ is the Eddington accretion rate (with proton mass $m_p$ and Thomson cross section $\sigma_T$).  The dependence of $P_{eq}$ on magnetic field, due to pressure balance at the Alfven radius, defines a birth line or spin-up line in the $P$-$\dot P$ diagram (see Figure~\ref{fig:PPdot}). During the accretion phase, the disc is bright in X rays and the source is a low-mass X-ray binary (LMXB), but after accretion stops, the source becomes a rotation-powered pulsar, spinning down due to magnetic dipole torques. 
This evolution scenario was initially confirmed by the detection of millisecond periods in LMXBs \cite{Wijnands98}, but recently several LMXBs have been caught red-handed, transitioning multiple times from accretion-powered to rotation-powered sources \cite{Stappers14,Roy14}. We discuss these state changes in section~\ref{sec:BW}. 

\subsection{The unexpected \g-ray millisecond pulsar revolution}
\label{sec:MSPrevolution}

Even though the magnetic fields of MSPs are relatively low (see Figure~\ref{fig:PPdot}, their short periods produce high enough electric fields to accelerate particles to $\gamma$-ray energies \cite{Harding05,ZhangCheng03}.  However, it was thought that MSPs cannot produce rich pair cascades unless there are distortions of the surface magnetic fields \cite{HM11} (see section~\ref{sec:mag}), so many were expected to be pair-starved \cite{Harding02}. There was a possible detection of one object, J0218-4232, by EGRET prior to \Fermi \cite{Kuiper00}, but the significance was marginal. So, the discovery of numerous \g-ray MSPs came as a surprise, accompanied by other surprises in their electrodynamics.

Figure~\ref{fig:PPdot} shows that the new \g-ray MSPs tend to be on the faster and more energetic side of the MSP population, as do young \g-ray pulsars with respect to canonical, not-recycled pulsars, but they surprisingly represent a larger fraction of the known MSP population (45\% over the same range in $P$, \Pdot, and $B_{LC}$) than the percentage of young \g-ray pulsars among not-recycled ones  (23\%  over the same range in $P$, \Pdot, and $B_{LC}$). This difference can stem from a larger overlap between the radio and \g-ray beams in the case of MSPs because of the small size of their magnetospheres. All \g-ray MSPs detected so far indeed show radio pulsations (see Figure~\ref{fig:PPdot}). The MSPs further show, on average, larger lags between the radio and \g-ray pulses, and a larger fraction of radio pulses that are aligned in phase with the \g-ray pulses than for the young pulsars.  These trends are expected if the radio emission comes from a larger radius relative to the light cylinder in the smaller MSP magnetospheres, where stronger aberration causes larger radio lags and even radio caustics that align with the \g-ray caustics at the highest altitudes \cite{Venter12}. 

As a second surprise, the old, recycled neutron stars have turned out to be the most efficient \g-ray pulsars, with efficiencies often approaching or exceeding 10\% (see Figure~\ref{fig:lum_effi}). These efficiencies relate to the large magnetic field strengths pervading the outer regions of the compact MSP magnetospheres, which compare with the values found near the light cylinder of young pulsars (see Figure~\ref{fig:lum_effi}). 

As a third surprise, most MSP \g-ray light curves showed narrow peaks lagging the radio peak, basically indistinguishable from those of young pulsars (see Figure \ref{fig:MSP_LCs}), instead of the broad peaks aligned with the radio peaks predicted for pair starved models \cite{Frack05}. This finding suggested that the \g rays come from the outer magnetosphere and that MSPs must be generating pair cascades with high multiplicity to screen most of their magnetospheres.  Indeed, many \g-ray and radio light curves from MSPs are reasonably well fit with slot-gap and outer-gap models \cite{Venter09,Johnson14}, implying that these old pulsars are still capable of producing pair plasma and force-free conditions over much of their magnetosphere (see Section \ref{sec:mag}).  Only a small subset of objects show \g-ray peaks preceding the radio peaks in phase and these are in fact better fit with pair starved models \cite{Johnson14}.  A third subset of MSPs show $\gamma$-ray peaks that are aligned in phase with the radio peaks, implying that the radio emission comes from high altitude relative to the light cylinder, forming caustic peaks like the \g-ray emission \cite{Venter12}. 

The $\gamma$-ray spectra from MSPs are similar to those of young pulsars, except that their phase-averaged photon spectra are somewhat harder \cite{Abdo13_2PC}. A recent study of 25 of the brightest \g-ray MSPs reveals a spectral sequence with \Edot: the phase-averaged SED broadens and softens as \Edot increases \cite{Renault15a}. This may require the building up of an additional radiation component, besides curvature radiation, from the primary particles. The phase-resolved spectra of these 25 MSPs show that the brighter peak in a double light curve has a harder spectrum \cite{Renault15b}, and that on-peak and off-peak emissions evolve differently across the sample: 
the \g-ray flux in the peaks depends on $\sqrt{\dot{E}_{rot}}$, indicating more screening of $E_\parallel$, while the flux away from the peaks has a dependence closer to \Edot, expected with less screening \cite{Renault15a}. This would indicate that the magnetospheres contain both regions of higher and lower pair multiplicity.  

\subsection{Millisecond pulsar arachnology}  
\label{sec:BW}

As mentioned above, \Fermi has been very successful in discovering new MSP binary systems known as Black Widows (BWs), probably because the MSPs in these systems are energetic and radiate pulsed \g-rays with high efficiency. 
BWs are close binary systems, with orbital periods of hours, containing a rotation-powered
MSP and a compact companion having very low mass, $\sim 0.01 - 0.05M_\odot$. The companion
stars in BWs undergo intense heating of their atmospheres by the pulsar wind, which drives a
stellar wind and rapid mass loss from the companion star. The mass loss results is a slow ablation of the companion, possibly evaporating it completely, resulting in an isolated MSP.  Redbacks (RBs) are similar systems \cite{Roberts11},
except that the companions have somewhat higher masses, $\sim 0.1 - 0.4M\odot$.  The MSPs in
both types of system are typically energetic, with \Edot of order $10^{25-27}$~W.

Before the launch of \Fermi these systems were rare, with only three BWs and one
RB known. The large amount of material blown off from the companion stars absorbs
and scatters the radio pulsations from the MSPs, making them difficult to detect at radio
wavelengths. The radio searches toward \Fermi-LAT unidentified sources
\cite{Ray12} have discovered 15 new BWs and 9 new RBs to date, making a present total of 28 of these systems. Ten and three respectively pulse in \g rays.
A few cases were found through blind searches of the \g-ray pulsation, with the help of optical observations that first found a binary system and its orbital period toward the \Fermi-LAT source. The first was 0FGL J1311.9-3419 which showed a 93-day optical modulation \cite{Romani12} caused by the one-sided heating of the companion star which is tidally locked to always have one side toward the pulsar as it heats its atmosphere. Using the orbital period and its derivative, a computationally intensive \g-ray blind search was able to detect a MSP, J1311-3430 \cite{Pletsch12}. For a short time, this was thought to be the first radio-quiet \g-ray MSP, but the radio pulsations were eventually discovered after searching at several different frequencies \cite{Ray13}.  This source turned out to be a Black Widow with the shortest orbital period yet known, with the MSP and companion separated by only the Earth-Moon distance.  The pulsar J2339-0533 was discovered with the same method of first detecting the orbital period through optical observations \cite{RS11,Kong12}, followed by the discovery of a radio pulsar, followed by a successful \g-ray blind search \cite{Ray14}.  This source turned out to be a Redback with a massive MSP and a companion mass of $0.26 M_\odot$.

Intrabinary shocks may form in BW and RB systems, at the location where the pulsar-wind pressure balances the pressure in the wind of the companion star.  The pulsar-wind termination shock, as in pulsar wind nebulae, is a possible site of particle acceleration and high-energy radiation \cite{AT93,HG90}, modulated at the binary orbital period.  Orbital modulation of marginal significance has been reported for several of these systems in the \g-ray band \cite{Dubus15} and there have been  detections of X-ray emission modulated at the orbital periods in the BW systems B1957+20 \cite{Huang12}, J2215+5135 and J2256-1024 \cite{Gentile14}. The X-ray emission is most likely synchrotron emission from accelerated electrons and positrons \cite{AT93,Bogdanov11}, but inverse Compton emission at \g-ray energies from these particles scattering soft photons from the companion star would also be expected.  A possible contribution of shock accelerated leptons in BW and RB systems to the observed cosmic-ray positron excess \cite{Adriani13,Accardo14} has also been suggested \cite{Venter15}. 

\g-ray MSPs also shed light on the last stages of the spinning-up of neutron stars, and on the transition from accretion-powered to rotation-powered activity. The J1023+0038 pulsar provides a useful test bed at a distance of 1.37 kpc toward the \g-ray source 2FGLJ1023.6+0040: an accretion disc was observed in the optical in 2000--2001; the detection of radio pulses from a MSP followed in 2007, until the pulses disappeared in 2013, quenched or eclipsed by the ionized flow from a new X-ray accretion disc \cite{Bogdanov14,Li14,CotiZelati14}. The interaction between the pulsar magnetosphere/wind and the accretion disc is very dynamic, switching between different spectral states in X rays, with and without orbital modulation \cite{Li14}, and producing radio continuum emission and sporadic flares in the optical, UV, and X-rays, suggestive of propellor modes  \cite{Bogdanov14}. 
The \g-ray flux quintupled when the radio pulses disappeared \cite{Stappers14,Takata14}. \g-ray emission before the state change was likely powered by the pulsar, yet with only marginal detection of ms pulses \cite{Archibald13}. With the reappearance of the accretion disc, unpulsed X rays and \g rays can also arise from synchrotron and inverse Compton emissions at the shock front between the pulsar wind and the accretion flow \cite{Tavani97}. It will be important to establish the wind origin of the \g rays, as they would signal that the pulsar is enshrouded, but is still actively accelerating particles. It will also be important to test whether the large \g-ray flux, which represents 14\% of the pulsar rotational power \cite{Stappers14}, requires further particle acceleration by the shock itself or not. Another nearby study case has turned up with J1227-4853 \cite{Roy14}.

\subsection{Millisecond pulsars as physics laboratories}
\label{sec:MSPlab}

Millisecond pulsars are extremely stable rotators, have very few glitches and little timing noise that plague young pulsar spin measurements.  They are therefore extremely good clocks and timing of MSPs has found a variety of applications, including searches for gravitational waves, neutron-star mass and radius measurements, constraints on General Relativity, and study of the intervening interstellar medium.  

A collection of MSPs spread out over the sky can be used as a gravitational wave detector \cite{Hobbs08}.  By timing at least 20 MSPs to a precision of $< 100$ ns, it is possible to make a direct detection of the stochastic gravitational wave background from distant merging black holes \cite{CS12}.  The technique involves looking for a characteristic pattern in the correlations of timing residuals (the difference between the expected and actual arrival time of the radio pulses) between pairs of MSPs, as a function of the angle between them \cite{HD83}.  It is also possible to make direct detection of individual sources of gravitational waves, such as a merger of binary supermassive black holes in a nearby galaxy \cite{Lommen12}, formation of a supermassive black hole \cite{TB76} or a highly eccentric black hole binary \cite{EN07}.  Several Pulsar Timing Array collaborations in the US, Australia, and Europe are actively searching for such signals, and the discovery of so many new MSPs by \Fermi to add to the arrays has accelerated the timescale for detection.  At least 12 new MSPs discovered toward \Fermi sources are now part of the arrays and more are constantly being discovered.  Current limits on the gravitational wave background at nHz frequencies are beginning to put limits on cosmic strings and the black hole merger rate \cite{Demorest13}.

Precision timing of MSPs has enabled the best measurements of neutron star masses.   In nearly edge-on binaries, Shapiro delays \cite{Shapiro64} (gravitational time delays) have been measured at the orbital phase where the line-of-sight to the MSP grazes the compact companion star.  The Shapiro delay causes a characteristic cusp in the time residual at conjunction, and the mass of the companion can be measured from the shape of the cusp, enabling a complete solution of the binary orbital elements, including a precise measurement of the mass of the neutron star.  Shapiro delays have been detected from several MSPs, including J1614-2230 yielding a mass of $1.97 \pm 0.04 M_\odot$ which is one of the largest values \cite{Demorest10}, and J1949+3106 with mass $1.47 ^{+0.43}_{-0.31} M_\odot$ \cite{Deneva12}.  A number of MSPs have measured masses significantly larger than the canonical $1.4 M_\odot$, most likely due to accumulation of mass during their accretion and spin-up phase.  In particular J0751+1807, also a \Fermi \g-ray pulsar, has a mass of $2.1 \pm 0.2 M_\odot$ from measurement of orbital decay and a Shapiro delay \cite{Nice05}, and J0348+0432 has the most precisely measured mass of $2.01 \pm 0.04 M_\odot$ \cite{Antoniadis13}. The recently discovered black widow J1311-3430 discussed above has a measured mass above $2.1 M_\odot$ from optical studies \cite{Romani12}.  Most interestingly, such high masses approaching or exceeding $2 M_\odot$ are constraining the equation of state  of nuclear matter, ruling out the softer solutions and quark matter.  

As a complement to the mass measurements, MSP radii can be constrained by modeling their pulsed X-ray light curves.  The paths of thermal X-rays, thought to come from polar caps heated by relativistic particles from the magnetospheric gaps \cite{HM02}, are bent by the strong gravitational field of the neutron star.  The light curves are distorted by the light bending, which is a function of the stellar compactness ($M_{NS}/R_{NS}$) and with extreme compactness an observer can see the other side of the star.  Modeling the thermal X-ray light curve of J0437-4715 \cite{Bogdanov13}, with its mass already measured at $1.76 \pm 0.02 M_\odot$ by radio timing \cite{Verbiest08}, constrains its radius to $R_{NS} > 11.1$~km.  Similar modeling of J0030+0451 \cite{Bogdanov07} yielded a radius constraint of $R_{NS} > 10.4$~km.  A much more precise use of this technique is planned for the Neutron Star Interior Composition Explorer (NICER), due to launch in 2016 \cite{Gendreau12}.  NICER will be able to measure stellar radii of MSPs to within 5\%, providing the most stringent constraints yet on the equation of state.

Timing of MSPs also provides interesting constraints on General Relativity and other gravitational theories.  Neutron star-white dwarf systems in particular are useful for testing the strong equivalence principle (SEP) of General Relativity \cite{Freire12}, which states that gravity is purely geometrical and acts in the same way regardless of composition, velocity or location, as in the weak equivalence principle, but also extends this to self-gravitating bodies.  One consequence is that the motion of orbiting bodies with strong self-gravity should be the same as those with weak self-gravity.  Since General Relativity is thought to be the only theory that satisfies the SEP, finding violations of it would question a central tenet of General Relativity.  Current tests using MSP binaries have set the strongest limits on violation of the SEP in strong gravity and also Lorentz invariance violation \cite{Nice05,Gonzalez11,Horns15}.  Recently a triple system consisting of a MSP J0337+1715 and two white dwarfs will provide the strongest constraints yet on the SEP in a strong gravitational field \cite{Ransom14}. In this system, the MSP and one of the white dwarfs, with gravitational binding energies different by five orders of magnitude, are both falling in the gravitational field of the other white dwarf.   

\section{From vacuum to charge-filled magnetospheres}
\label{sec:mag}
\subsection{Early models and the vacuum magnetosphere}
\label{sec:vacuum}
Study of the spectrum and light curves of \g-ray pulsars began soon after their discovery.  As there were no quantitative global models of pulsar magnetospheres until 1999, early models for pulsar high-energy emission derived the location and particle acceleration properties assuming boundary conditions based on local physics.  The first models assumed that the acceleration and high-energy emission is located above the pulsar polar caps, with curvature radiation as the primary mechanism \cite{Harding78,MS79}.  Additional synchrotron radiation from electron-positron pair cascades (see below) was later also considered \cite{DH96}, and predicted sharp high-energy cutoffs in the spectrum due to magnetic pair production attenuation at a few GeV.  As discussed above, such super exponential cutoffs have not been observed by \Fermi-LAT, indicating that the \g-ray emission is not coming from the polar cap.  Models for high-energy emission from the outer magnetosphere had in fact already been developed.  
The outer gap model \cite{Cheng86,RY95} locates the acceleration region above the null charge surface (the surface where the corotation charge for a static dipole changes sign) near the last open field lines.  The slot gap \cite{Arons83,MH04} is an extension of the polar cap where acceleration occurs in a narrow charge-depleted zone along the last open field lines from the neutron star surface to near the light cylinder.  Both of these models can account for the variety and major characteristics of the \Fermi pulsar light curves, but do not provide enough power to account for the observed luminosities \cite{Pierbattista12}.  Furthermore, neither of these models is consistent with the global magnetosphere current closure.

Earlier emission models all used the retarded vacuum solution \cite{Deutsch55} which was the only one available. In this solution, a quadrupolar electric field is induced by the rotation of the neutron star and the dipole magnetic lines sweep back near and beyond the light cylinder into a toroidal pattern, becoming a vacuum electromagnetic wave at infinity.  However, real pulsar magnetospheres are filled with charges and currents and are far from vacuum. 
There has been recent significant progress toward constructing simulations of the global pulsar magnetosphere, although early attempts started soon after the discovery of pulsars over forty year ago.  Over these years, there have been two computational approaches to solving the problem of how the pulsar magnetosphere fills with charges. 
One approach, in MHD, is to solve for the self-consistent electromagnetic fields, current and charge densities of a rotating star with a dipole moment.  The second approach is to compute the individual particle response to the electromagnetic fields, through N-body simulations, or through particle-in-cell (PIC) simulations that also compute the self-consistent change in the fields due to the particles.  

\subsection{MHD magnetosphere models}
\label{sec:magmodels}
In the MHD approach, the equation for the aligned force-free magnetosphere was formulated by Michel in 1973 \cite{Michel73}, but not solved until 1999 \cite{CKF99}
by numerical computation.  This solution showed that a current sheet forms along the spin equator where magnetic field lines from opposite poles come together and also showed, for the first time, how the magnetospheric current closes, flowing out of the polar regions and returning through the current sheet along the last open field lines.  \cite{Timokhin06}.  The time-dependent solution for the oblique magnetosphere (Figure \ref{fig:FF}) appeared in 2006 \cite{Spitkovsky06}.  These simulations solve the time-dependent Maxwell's Equations,  in the limit of inertia-free plasma and under the imposed condition that ${\bf E \cdot B} = 0$, with a current along each field line that is adjusted to sustain the force-free electric field. 
Although the force-free solutions contain the distribution of charges and currents, including current closure, in the magnetosphere, they demand, by definition, that the electric field along the magnetic field direction vanish (ideal MHD). These force-free models can provide the global configuration of the pulsar magnetosphere but can tell us little about particle acceleration and emission. 

In the last few years, two groups \cite{Kala12a,Li12} have begun developing dissipative pulsar magnetosphere models, where the ideal MHD condition for zero parallel electric field is dropped.  These models assume a formulation for Ohm's law relating the macroscopic current to the electric field along magnetic field lines through a macroscopic conductivity parameter.  As a result they obtain the distribution of $E_\parallel$, allowing particle acceleration and radiation.  The dissipative magnetospheres have been used to model the high-energy emission from pulsars \cite{Kala12b,Kala14,Brambilla14}, showing that the acceleration and emission is widely distributed over the magnetosphere, concentrating in lobes over the polar cap for small conductivity, but progressively moving to the open field boundary and out to the current sheet beyond the light cylinder for high conductivity.  The predicted characteristics of both the \g-ray light curves and phase-resolved spectra generally agree with those of young pulsars detected by \Fermi-LAT, in the case of the highest conductivity (Figure \ref{fig:FIDO}).  
Although these dissipative models formally include accelerating electric field, the conductivity is introduced ad hoc, based on empirical requirements to fit specific pulsar light curves rather than on considerations of  microphysics. As successful as the dissipative MHD pulsar magnetosphere simulations have been, they also lack fundamental self-consistency.  They assume a free and abundant supply of whatever charges are needed to support the macroscopic currents, without specifying the sources of the charges.  Furthermore, the MHD codes derive only the macroscopic currents and charge densities, without knowledge of the type or contribution of various charge carriers making up the currents.  

\begin{figure}[!t]
  \centering
  \includegraphics[width=7.0cm]{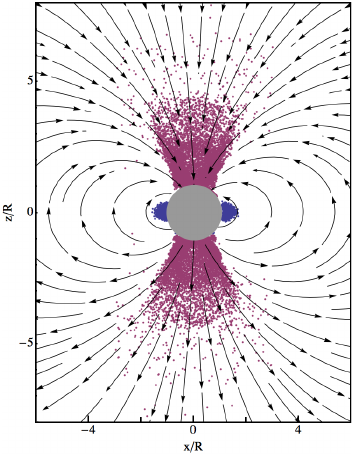} \includegraphics[width=7.0cm]{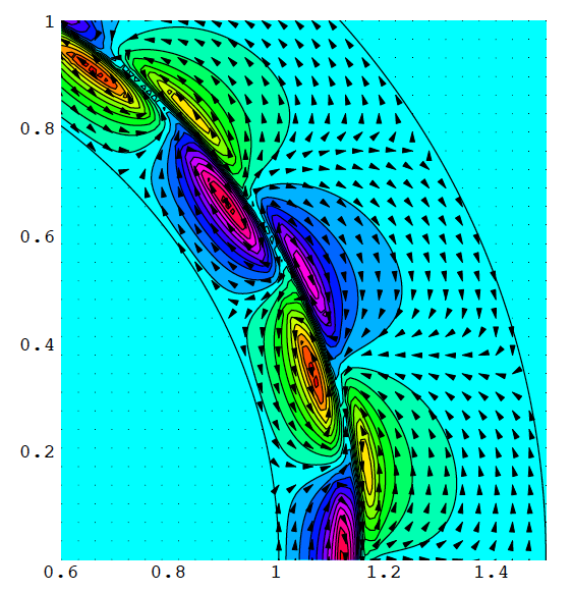}
  \caption{\textit{Left:} slice through a 3D particle simulation of an aligned rotator \cite{PS14} where a charge-separated electrosphere develops. Arrows are magnetic field lines and color is sign of charge: red for positrons, blue for electrons. \textit{Right:} diocotron instability in the equatorial plane of a charge-separated aligned rotator \cite{Petri02b}, showing density perturbations (color) and velocity field (arrows).}
  \label{fig:esphere}
\end{figure}

\subsection{N-body and Particle-in-cell magnetosphere models}
\label{sec:simul}
In this sense, the particle approach is superior in principal and better suited to constructing a fully self-consistent pulsar magnetosphere.  The challenge is computational, since  particle codes are considerably more demanding than MHD codes.  For this reason, progress on this front has been slower until very recently.  The first attempt to simulate how an aligned  vacuum pulsar magnetosphere fills with charge \cite{KM85}, assumed to be supplied only from the neutron star, found that charges are pulled from the surface and arrange themselves in an equilibrium configuration with a dome over the polar cap and a torus of opposite charge sign along the equator, separated by vacuum.  Such an ``electrosphere" (see Figure \ref{fig:esphere}a) would be inactive, with no currents or wind formation.  But the torus was later found to develop a diocotron instability (see Figure \ref{fig:esphere}b)\cite{Petri02a,Petri02b}, which led to cross field diffusion of the particles and partial filling of the magnetosphere, but not out to the light cylinder \cite{Petri03}.  A PIC simulation of the electrosphere \cite{SA02} showed that the diocotron instability developed even faster for an inclined rotator, filling the magnetosphere out to the light cylinder, but for computational reasons the simulation was limited in time and number of particles.  

\subsection{Pair cascades}
\label{sec:cascades}
\begin{figure}[!t]
  \centering
  \vskip -1.0cm
  \includegraphics[width=14cm]{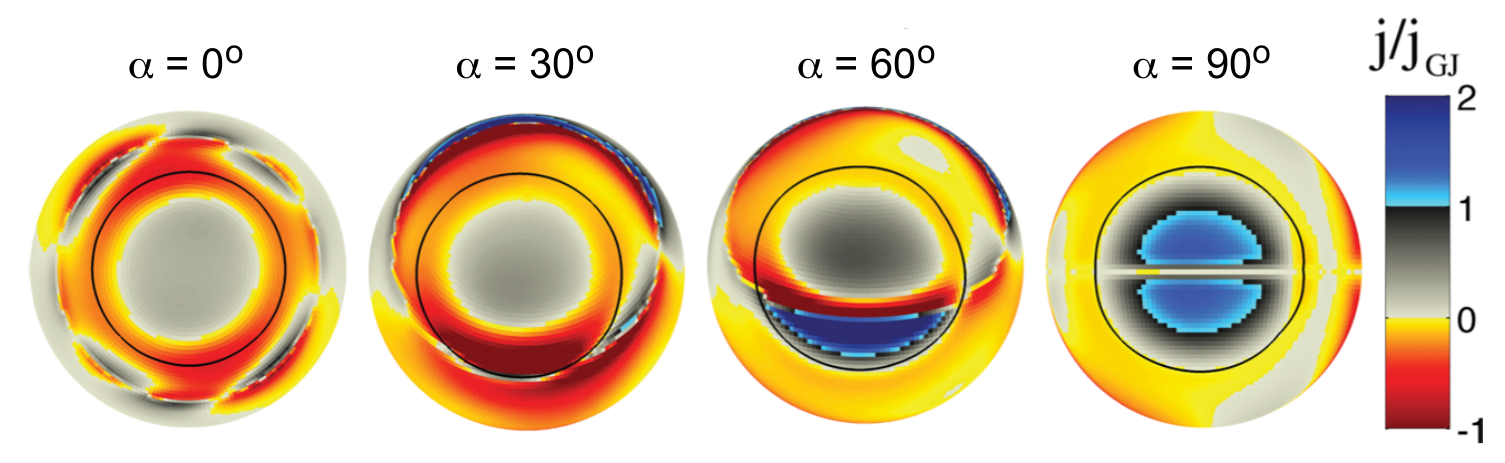}
  \caption{Pplar view of the current density distributions (current density $j$ relative to the Goldreich-Julian current density $j_{\rm GJ}$) on the polar cap in a force-free magnetosphere for different pulsar inclination angles $\alpha_B$.  The black curves outline the polar cap rim (boundary between the open and closed field lines). Polar cap pair cascades do not take place on field lines where $0 < j/j_{\rm GJ} < 1$ (gray areas), but are active on field lines where $j/j_{\rm GJ} > 1$ (blue) and $j/j_{\rm GJ} < 0$ (yellow/red).  Adapted from \cite{TA13}.}
  \label{fig:PairCascades}
\end{figure}

In a real pulsar magnetosphere, the source of charges is not only particles from the neutron star but also pair cascades, which can occur both near the polar caps and in the outer gaps.  The microphysics of these electron-positron pair cascades, first proposed by Sturrock \cite{Sturrock71}, has been studied in detail over the last thirty years  
and is widely considered to be critical for supplying charges to the magnetosphere as well as plasma for the observed coherent radio emission.  The electromagnetic cascades above the polar caps \cite{DH82}  can efficiently supply charges through $\gamma$-rays that undergo conversion to electron-positron pairs by the strong magnetic field, $\gamma + B \rightarrow e^{\pm}$, for photons with pitch angle $\theta$ and energy $E_{\gamma} \gtrsim 2 m_e c^2 \left(\frac{4.4\,10^9\,\rm{T}}{B\sin \theta}\right)$  \cite{Erber66,DH83}. These cascades are initiated by the acceleration of primary electrons in strong electric fields above the polar-cap surface.  The electric fields arise from a deficiency of charge compared to that needed to screen the electric field parallel to the magnetic field.  For surface $B < 10^{10}$~T, charges of either sign can be pulled from the surface whose temperature (which has been measured to be in the range $T \sim 10^5-10^6$ K) exceeds the ion and electron
thermal emission temperatures \cite{ML07}.  Although $E_\parallel = 0$ at the neutron-star surface, the space charge along open field lines above the surface falls short
of the corotation charge, due to the curvature of the field \cite{AS79} and to general relativistic inertial frame dragging \cite{MT92}.  The $E_\parallel \ne 0$  generated by the charge deficit accelerates particles, which radiate inverse Compton (IC) photons (at particle Lorentz factors  $\sim 10^2 - 10^6$) and curvature photons (at Lorentz factors  $\gsim 10^6$).  Curvature and IC photons from the particles reaches tens of GeV, creating pairs in excited Landau states. The pairs lose their perpendicular momentum by emitting synchrotron photons that create more pairs. These steady-state cascades yield pair multiplicities up to $10^4$ \cite{HM11} that are a strong function of pulsar period and surface magnetic-field strength.  However, recent studies of time-dependent polar-cap pair cascades \cite{Timokhin10,TA13} have found that the cascades are only steady-state for primary currents that are finely tuned to the Goldreich-Julian current, $J_{\rm GJ} = \rho_{\rm GJ} c$, where $ \rho_{\rm GJ} = 2\epsilon_0 \Omega B$ is the charge density necessary to screen $E_\parallel$.  Since the global MHD models find a range of currents across the polar cap, the pair cascades will in most cases be time variable.  Since the time-variable gaps have somewhat higher $E_\parallel \ne 0$, the pair multiplicities can reach $10^5$ \cite{TH14}.  However, pair cascades are not active for all values of current required by the global models, particularly for the range $0 < J/J_{\rm GJ} < 1$ (see Figure \ref{fig:PairCascades}) that exists over most of the polar cap for smaller inclination angles.

Pair creation is also expected to occur in outer gaps, if they exist in certain magnetospheric configurations.  In this case, the pairs are created by the photon-photon process
in which $\gamma$-ray photons of energy $\epsilon_\gamma$ from primary particles accelerated in the gap and X-ray photons $\epsilon_X$ from secondary particles or from the hot polar caps produce an electron and positron \cite{Cheng86,ZhangCheng97}.  The threshold for this process is $\epsilon_\gamma \epsilon_X = 2 / (1 - \cos\theta)$, where the energies are in units of $mc^2$ and $\theta$ is the angle between the photon momenta.  Due to the low density of thermal X-ray photons in the outer magnetosphere, the pair multiplicity of outer-gap cascades is lower than that of polar cap cascades.   The pairs are born with non-zero pitch angles and radiate synchrotron emission from optical to hard X-ray wavelengths, initiating the cascade. The pairs and primaries also undergo inverse Compton scattering with soft synchrotron and curvature photons \cite{Tang08,Hirotani08}.

\subsection{Magnetosphere models with pairs}
\label{sec:pairs}
Recently, several groups have performed particle simulations of a global pulsar magnetosphere with simplified pair injection.  Yuki \& Shibata \cite{YS12} performed an N-body simulation of an aligned pulsar magnetosphere, injecting a low multiplicity of pairs at locations where the electric field is above a given threshold.  They find that both outer gaps and slot gaps are simultaneously present below and above the neutral current zone of the open field region where currents flow in opposite directions.
Very recently, several groups have built PIC simulations of a pulsar magnetosphere.  3D PIC simulations of an aligned rotator \cite{PS14} were able to reproduce the ideal force-free magnetosphere, assuming injection of sufficient neutral pair plasma everywhere in the magnetosphere.  Chen \& Beloborodov \cite{CB14} explored more limited, but highly simplified, pair injection to a PIC simulation of an aligned rotator. They found that when pairs are injected throughout the magnetosphere, a force-free configuration results with instabilities developing in the current sheet. If pairs are injected only near the polar caps and when emitted curvature photons reach a threshold energy, they find that the system relaxes close to the electrosphere solution (see section \ref{sec:simul} and Figure \ref{fig:esphere}). However, the aligned rotator is a pathological case since, as discussed above, current in the range $0 < J/J_{\rm GJ} < 1$ exists over most of the polar cap which quenches the pair cascades.  Philippov et al. \cite{Philippov14} presented a PIC simulation of an oblique pulsar magnetosphere for both cases where pairs are injected everywhere, reproducing the force-free solutions, and where pairs are injected when primary particles reach a given threshold energy.  In the limited pair injection case, they find that as the inclination angle increases, the regions of pair creation expand to fill most of the open field region and plasma with high multiplicity sustains the pulsar wind.  While these results are very encouraging, simulations with more realistic pair injection from full cascade calculations will be required to fully describe the geometry and characteristics of the acceleration regions in the magnetosphere.

\subsection{From the polar cap to the current sheet}
\label{sec:sheet}
As a result of the recent progress in defining and understanding the global pulsar magnetosphere both inside and outside the light cylinder, models for the production of high-energy emission near or in the current sheet have gained favor.  Thus over the last forty years, the favored site of pulsar high-energy emission has migrated from near the polar caps, to the outer magnetosphere, to outside the light cylinder.  Using the vacuum magnetic-field solution, the velocity of particles moving along the field would exceed the speed of light beyond the light cylinder, so emission models using this solution were restricted to less than $R_{\rm LC}$.  But the solution of the force-free and dissipative magnetic-field structure has made it possible to solve the dynamics and radiation of particles outside the light cylinder in the inertial observer's frame (e.g. \cite{BS10}).  However  the actual location and mechanism for the emission remain unclear.  Global dissipative MHD models find that acceleration of particles near, but not in, the current sheet by electric fields induced by the rotation can produce the observed $\gamma$-ray emission through curvature radiation \cite{Kala14,Brambilla14}.  It has also been proposed that reconnection in the current sheet near the light cylinder can energize particles that then radiate via synchrotron emission \cite{Lyubarski96,Petri11} up to possibly 100 GeV energies \cite{US14}, if Doppler boosting due to particle outflow is taken into account.  However, these models make the simplified assumption of equipartition between electromagnetic and particle energy density, resulting in very high plasma temperatures.  This would be problematic for young pulsars, whose observed $\gamma$-ray efficiencies are less than a few percent of spin-down luminosity.  In addition, they assume neutral current sheet solutions, where the real pulsar current sheet is charged and this changes the solution significantly \cite{DeVore15}.

\section{From permanence to impetuousness}
\label{sec:imp}
\subsection{Pulsar variability}
\label{sec:variab}

The large sensitivity of radio telescopes and their long-term monitoring of pulsars has enabled variability studies of the radio pulses over a wide range of time scales, from seconds to years. Many variable phenomena have been found \cite{Lyne10}. Pulsar ``nulling'' corresponds to the abrupt disappearance and sudden reappearance of the pulses after a few stellar rotations or a few hours. In extreme nulling cases, the on and off intervals can last weeks or months. Such intermittency occurs quasi-periodically. Some pulsars also switch back and forth every few minutes between two, occasionally three, modes with well-defined, but different mean pulse profiles. Further descriptions and examples are reviewed in \cite{Keane13} and chapters of \cite{Becker09}.

Nulling events imply persistent changes in the charge and current distributions along open field lines, possibly between charged-filled, nearly force-free conditions in the ``on'' state and pair-starved, nearly vacuum conditions in the ``off'' state \cite{Li12}. Changes in the magnetospheric currents concur with changes in the braking torque on the star, as evidenced in correlated variations in the radio pulses and in \Pdot  \cite{Lyne10}. The spin-down rate is indeed larger by a factor $\frac{3}{2}(1 + \sin^2 \alpha_B)/ \sin^2 \alpha_B$ for a force-free magnetosphere than for a vacuum dipole \cite{Spitkovsky06,Petri12}, so changes in the charge supply can impact the rotation, even if a dense plasma remains trapped in the closed-field region \cite{Li12}. Mode switching, on the other hand, suggests current reorganizations from some open field lines to others, or the shutting off of specific components in composite, multi-beam light curves.  

Isolated neutron stars do not smoothly slow down. Shear stress between the magnetic flux tubes moving inside the neutron star, but anchored in the solid crust, can cause the unpinning of the rotational vortices from the crust. These ``glitch'' events are detected as abrupt spin-up jumps in otherwise spinning-down stars and they happen in neutron stars younger than a few million years of age \cite{Espinoza11}.
Several \g-ray pulsars have experienced glitches during the \Fermi survey \cite{Ray12}, but they have not triggered any detectable \g-ray variability \cite{Acero15_3FGL}.

\begin{figure}[!t]
\centering
\includegraphics[width=\textwidth]{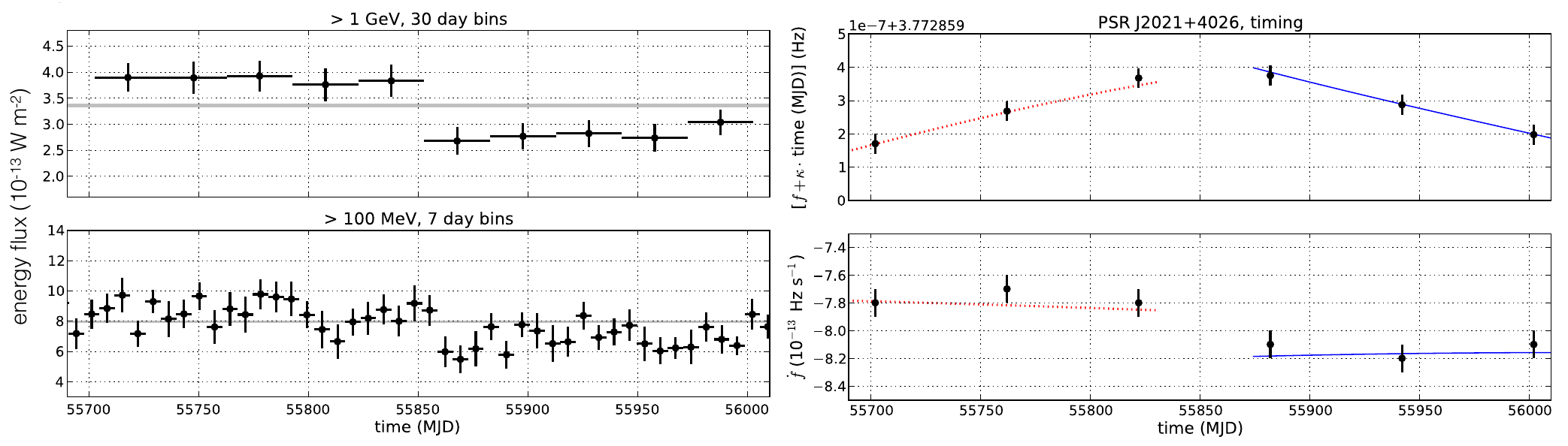}
\caption{Time evolution over 300 days of the \g-ray flux in different time bins and energy bands (left) and of the rotational frequency $f$ and its time derivative $\dot{f}$ (right) of the mode-switching pulsar J2021+4026  \cite{Allafort13_PSR2021}.}
\label{fig:PSR2021}
\end{figure}

Global dynamic changes in the magnetosphere are very intriguing and complementary information can be gained across the electromagnetic spectrum: the radio pulses respond to the microphysics of the complex emission mechanism, but they can also probe rapid macroscopic changes in the magnetosphere down to scales of seconds; X rays can probe variations in particle streams bombarding the polar caps; \g rays can probe global changes in the current circulation and in the level of pair production, although in time intervals longer than typically a week. The pulsar B0943+10 offers an important example of simultaneous switching of the radio and X-ray emission  as we look down onto the polar cap of this nearly aligned rotator ($\alpha_B \simeq 15^{\circ}, \zeta_{obs} \simeq 9^{\circ}$): the radio pulse changes in brightness and shape as an additional thermal X-ray hot spot turns on and off  \cite{Hermsen13}. These observations imply a rather globald magnetospheric change within a few seconds, which challenges current theories. The radio-quiet pulsar J2021+4026 has recently exposed another type of mode change in \g rays: its rotational power, \g-ray flux, and pulse profile all changed stepwise within a week (see Figure~\ref{fig:PSR2021}) \cite{Allafort13_PSR2021}. The flux had slightly, but steadily increased during the three years preceding the jump. We are probably viewing this pulsar at large angle from the magnetic axis, with $|\alpha_B - \zeta_{obs}|$ of order 60\degr--80\degr \cite{Pierbattista14}. This mode change resembles \g-ray changes recorded in Vela in 1975 and 1981 \cite{Grenier88}. In both objects, one peak has remained stable through the switching (1st peak for Vela, 2nd for J2021+4026), thus invalidating possible changes in the telescope sensitivity and data reconstruction. In both objects, the emission changed mostly outside the caustic peaks, in particular in the third peak, and at low energies, below the apex of the energy spectrum which peaks near 1 GeV. These mode changes, however, differ in their time profile:  J2021+4026 went through a step change whereas Vela experienced a jump and recovery phase with a rise time of $11.4\,{\pm}\,1.0$ days and a decay time of $137\,{\pm}\,39$ days. Whether these \g-ray mode changes signal a global magnetospheric current reconfiguration or only a partial one in the production zone of the third peak and bridge emission poses a stimulating problem. Additional examples of \g-ray mode changes are eagerly awaited to probe magnetospheric electrodynamics. 

\subsection{Flares from the Crab nebula}
\label{sec:Crab}

\begin{figure}[!t]
  \centering
  \vskip -1.0cm
  \includegraphics[width=14cm]{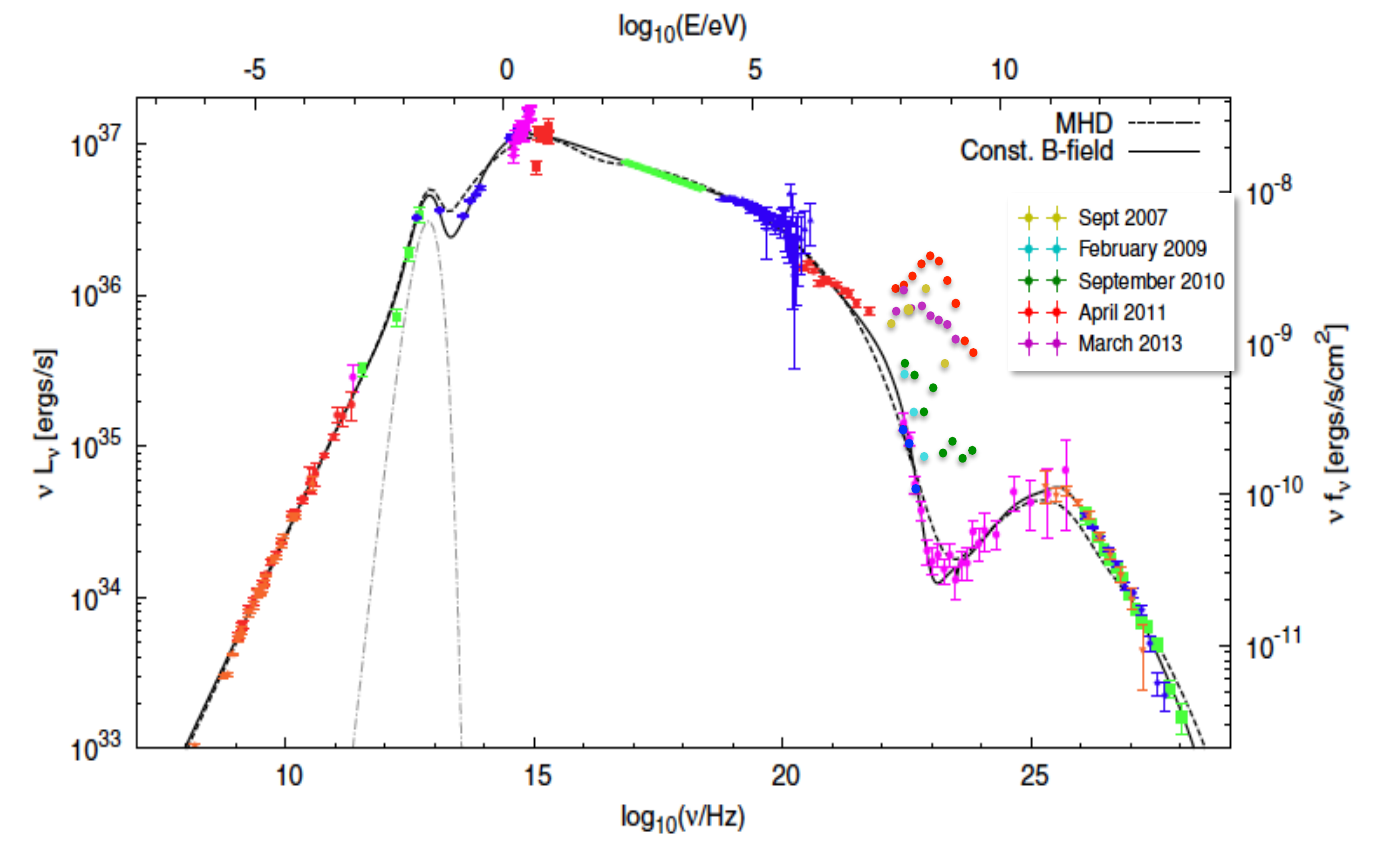}
  \caption{Spectral energy distribution of the Crab nebula from the radio to very high energy \g rays (from \cite{Meyer10}).  Also shown are the detected flares above 100 MeV, from  \cite{Buehler14}.}
  \label{fig:CrabFlares}
\end{figure}

Most of the spin-down power output of a pulsar appears as a magnetized wind of electrons and positrons, confined by the supernova shell and visible as a synchrotron nebula (see \cite{Hewitt15}). 
The Crab nebula has for decades been the standard candle for flux calibration, 
widely used as a flux unit in X-rays and \g rays.   Given its assumed constancy, observations of bright \g-ray flares from the Crab, first by AGILE \cite{Tavani11} and then by \Fermi-LAT \cite{Abdo11_CrabFlare} was a complete surprise.  The Crab flux above 100 MeV increased by factors of several for periods of a few days in October 2007, August 2008, February 2009 and September 2010.  This variation was only seen in the lower energy bands up to a few hundred MeV, thought to be the synchrotron component of the nebular emission (see Figure \ref{fig:CrabFlares}).  During these flares, the pulsed flux was constant so the flares are from the nebula. The largest flare to date was observed by \Fermi-LAT in April 2011 \cite{Buehler12}, when the flux rose for about six days over the quiescent flux level by a factor of 30 at peak.  Fast variability of less than one hour was seen, implying a very compact region ($< 0.04"$), and a new spectral component peaking at 500 MeV and extending to 1 GeV appeared.  As dramatic as this flare was at $\gamma$-ray energies, it represented only 1\% of the pulsar spin-down power and no flares were observed at optical or X-ray \cite{Weisskopf13} wavelengths.  Somewhat smaller flares with flux peaking at a level 16 times higher than quiescent levels were observed by \Fermi-LAT in March and October 2013 \cite{Mayer13}.  Simultaneous observations at TeV energies with HESS and VERITAS \cite{Aliu14,HESS14} revealed no flux increases.  

The Crab flares were not only surprising for breaking the paradigm of flux constancy, but are surprising and puzzling also for exceeding the standard upper limit of synchrotron radiation energy from accelerated electrons.  This limit comes from the maximum energy achievable through acceleration of electrons (or positrons) that is limited by synchrotron radiation losses \cite{Guilbert83,DeJager96}.   Balancing the fastest acceleration rate, the gyrofrequency ($\sim eB/mc$), with the synchrotron loss rate ($\propto \gamma^2\,B^{2}$), where $B$ is the magnetic field and $\gamma$ is the particle energy,  the maximum synchrotron radiation energy, $E_{\rm syn}^{\rm max} \simeq 9\,mc^2/4\alpha_f = 160$ MeV, where $\alpha_f$ is the fine structure constant, is independent of all source parameters.  The peak of the April 2011 flare spectrum at 500 MeV clearly violates this limit.  However, this limit assumes that the electric field $E < B$ in the region where the particles are accelerating, a condition that holds for induction $E$ fields in shocks and pulsar magnetospheres.  But $E > B$ can occur in a reconnection layer where $B \sim 0$ \cite{Uzdensky11}.  The current sheet (see Section \ref{sec:mag}) that separates field lines of opposite polarity in the pulsar wind is a likely site for reconnection \cite{Coroniti90}.

Synchrotron self-Compton models \cite{DJH92,AA96} provide a very good description of the quiescent emission of the Crab nebula.  Fits to the observed data indicate that the emission from radio to around 100 MeV is synchrotron radiation from a broad spectrum of electrons (and positrons) and the emission at higher energies is inverse Compton scattering of the synchrotron photons by these same electrons.
The traditional model for particle acceleration in the Crab nebula centers on the pulsar wind termination shock \cite{KC84}, where balance between the wind pressure and the nebular energy density confined by the (unseen) supernova shell locates the shock at a distance of $\sim 3 \times 10^{15}$~m \cite{RG74}.  The acceleration mechanism near pulsar wind termination shocks is not understood, but is known to be highly efficient since the bolometric luminosity of the Crab nebula is about 20\% of the pulsar spin-down luminosity and the inferred maximum particle energy, $\sim 10^{16}$ eV, is at least 10\% of the available voltage across open field lines \cite{DeJager96}.  

First-order Fermi acceleration of the particles at the termination shock is expected to be inefficient because of the relativistic speed of the shock (few particles can cross back upstream of the shock) and because of the nearly toroidal geometry of the magnetic field in the spin equatorial plane: the field lines being parallel to the shock hamper particle diffusion across the shock. 
The diffusive first-order Fermi mechanism becomes problematic unless most of the magnetic energy is converted to particle energy upstream of the shock \cite{Sironi09}.  Since the ratio of magnetic energy to particle energy, $\sigma \gg 1$ at the pulsar light cylinder, an unsolved mystery known as the ``$\sigma$ problem" is how $\sigma \ll 1$ occurs before the wind reaches the termination shock to enable efficient particle acceleration there.  A suggested alternative to Fermi acceleration, resonant absorption of ion cyclotron waves \cite{Hosh92}, has the problematic requirement that most of the pulsar spin-down energy be in ions upstream of the shock. Reconnection in the current sheet of the striped wind has been suggested as a mechanism for transferring energy from the magnetic field to particles, but the rate of reconnection is likely to be too low to produce $\sigma \ll 1$ upstream of the termination shock \cite{LK01} unless compression by the shock can drive faster reconnection \cite{PL07,Sironi11}.  

Since traditional models of acceleration at the termination shock can only give the maximum synchrotron photon energy of 160 MeV, alternative models for the Crab flares have been suggested. Reconnection \cite{Cerutti12,Cerutti13} in a thin layer where $B \rightarrow 0$ can avoid the classical synchrotron loss limits and allow electrons to accelerate up to energies of a few PeV before diffusing out of the reconnection layer and radiating synchrotron radiation up to 0.5 - 1 GeV as observed in the April 2011 flare.  Doppler boosted synchrotron emission from the termination shock was proposed to increase the synchrotron photon energy to 0.5 GeV and to produce the variability in the flares \cite{KL11,Lyutikov12}.  However, one would then also expect variability at all wavelengths which has not been observed.  MHD simulations \cite{KL03} have shown that the termination shock has an arch-like shape and comes much closer to the pulsar near the spin axis than it does near the spin equator.  The proposed site of the flares in this model, the inner knot where the termination shock is near the base of the jet, has generated recent interest when high-resolution near infrared images revealed activity correlated with the March 2013 flare \cite{RMW14}.  This could in fact be the site of the reconnection that produces the flares.

\section*{Acknowledgements}
We thank the pulsar timing and search consortia for their key support of the \Fermi-LAT pulsar analyses and we thank the \Fermi-LAT pulsar team and the pulsar theory group at NASA GSFC for many inspiring discussions.
I. G. acknowledges research support from the Institut Universitaire de France. AKH acknowledges support from the NASA Astrophysics Theory Program. 



\end{document}